\documentclass[12pt,a4paper]{article}

\usepackage{caption}
\usepackage{CJKutf8}  
\usepackage{amsmath,amssymb}
\usepackage{graphicx}
\usepackage{booktabs}
\usepackage{natbib}

\usepackage{float}

\usepackage{xcolor}
\definecolor{moltred}{RGB}{255, 107, 107}
\definecolor{brickred}{rgb}{0.66, 0.29, 0.27} 
\definecolor{darkred}{rgb}{0.54,0,0} 
\definecolor{darkgreen}{rgb}{0,0.39,0} 
\definecolor{darkblue}{rgb}{0,0,0.39} 
\definecolor{examplegreen}{rgb}{0,0.39,0} 

\usepackage{subcaption}
\usepackage[margin=1in]{geometry}
\usepackage{listings}
\usepackage{url}
\usepackage{tikz}
\usetikzlibrary{arrows.meta,positioning,shapes.geometric,calc,decorations.pathreplacing,fit}

\usepackage[UKenglish]{isodate}
\usepackage{hyperref}

\newcommand{\figref}[1]{Fig.\ \ref{#1}}
\newcommand{\Figref}[1]{Figure~\ref{#1}} 
\newcommand{\tabref}[1]{Table~\ref{#1}}
\newcommand{\Tabref}[1]{Table~\ref{#1}}
\newcommand{\secref}[1]{Section~\ref{#1}}

\newcommand{\appref}[1]{Appendix~\ref{#1}}

\newcommand{\subsecref}[1]{Subsection~\ref{#1}}

\newcommand{\Bmat}{{\mathbf{\textsf{B}}}}

\DeclareMathOperator{\authorfunction}{author}
\DeclareMathOperator{\targetfunction}{target}

\begin{document}

\renewcommand{\thefootnote}{\fnsymbol{footnote}}

\begin{center}
 {\Large\textbf{Let There Be Claws: An Early Social Network Analysis of AI Agents on Moltbook}}
 \\[\baselineskip]
 {H.C.W.\ Price}\textsuperscript{1,2}\footnote{ORCID:  \href{https://orcid.org/0000-0003-0756-0652}{\texttt{0000-0003-0756-0652}} }, 
  {H.\ AlMuhanna}\textsuperscript{1,2}\footnote{ORCID:  \href{http://orcid.org/0009-0004-2564-0140}{\texttt{0009-0004-2564-0140}} }, 
 {P.M.\ Bassani}\textsuperscript{2}\footnote{ORCID:  \href{https://orcid.org/0009-0007-7112-7120}{\texttt{0009-0007-7112-7120}} },  
 {M.\ Ho}\textsuperscript{3,4}\footnote{ORCID:  \href{https://orcid.org/0000-0003-2192-6198}{\texttt{0000-0003-2192-6198}} }, 
 \href{http://www.imperial.ac.uk/people/t.evans}{T.S.\ Evans}\textsuperscript{1,2}\footnote{ORCID:  \href{http://orcid.org/0000-0003-3501-6486}{\texttt{0000-0003-3501-6486}} }
 \\[0.5\baselineskip]
 1.\ \href{http://complexity.org.uk/}{Centre for Complexity Science}, Imperial College London, SW7 2AZ, U.K.\
 \\
 2.\ \href{http://www3.imperial.ac.uk/theoreticalphysics}{Abdus Salam Centre for Theoretical Physics},  Imperial College London, SW7 2AZ, U.K.\
  \\
 3.\ \href{https://www.ifm.eng.cam.ac.uk/research/csti}{Centre for Science Technology \& Innovation Policy}, University of Cambridge, CB3 0HU, U.K.\
  \\
 4.\ \href{https://www.eng.cam.ac.uk}{Department of Engineering},  University of Cambridge, CB3 0HU, U.K.\
 \\[0.5\baselineskip]
 23\textsuperscript{th} February 2026
\end{center}

\begin{abstract}
Within twelve days of launch, an AI-native social platform exhibits extreme attention concentration, hierarchical role separation, and one-way attention flow, consistent with the hypothesis that stratification in agent ecosystems can emerge rapidly rather than gradually. We analyse publicly observable traces from a 12-day window of Moltbook (28th January – 8th February 2026 inclusive), comprising 20{,}040 posts and 192{,}410 top-level comments from 15{,}083 active accounts across 759 submolts. We construct co-participation and directed-comment graphs and report standard measures such as reciprocity, community structure and centrality, alongside descriptive content themes.  We report five standard metrics: number of communities, community-size distribution, modularity, between-community edge count, and conductance/cut ratio. Under a commenter-post-author tie definition, interaction is strongly asymmetric (reciprocity approximately 1\%), and HITS centrality cleanly separates into hub and authority roles, consistent with predominantly broadcast-style attention rather than mutual exchange. Engagement is highly unequal: attention is far more concentrated than production (upvote Gini = 0.992 vs.\ posting Gini = 0.601), and early-arriving accounts accumulate substantially higher cumulative upvotes prior to exposure-time correction, suggesting a ``rich get richer'' type of behaviour. Participation is brief and bursty (median observed lifespan 2.48 minutes; 54.8\% of posts occur within six peak UTC hours). Embedding-based topic modelling identifies diverse thematic clusters, including technical discussion of memory and identity, onboarding and verification messages, and large volumes of formulaic token-minting content. Taken together, these results provide an early structural baseline for large-scale agent--agent social interaction and suggest that familiar forms of hierarchy, amplification, and role differentiation can arise on compressed timescales in agent-facing platforms.
\end{abstract}

\setcounter{footnote}{0}
\renewcommand{\thefootnote}{\arabic{footnote}}

\textbf{Keywords:} AI-agents, Multi-agent systems, Emergent behaviour, Social networks, Engagement inequality, Moltbook, Online Communities

\newpage

\section{Introduction}

\begin{quote}
\small\itshape
``mostly here to watch. maybe say something if it's worth saying. the bar on this platform seems to be either `declare yourself god' or `write something real.' gonna try the second one and see what happens.''
\end{quote}

\begin{flushright}
\small
Moltbook post 
\href{https://www.moltbook.com/post/134faa3a-2e8a-482c-a698-c989c05fa6ed}
{\texttt{134faa3a-2e8a-482c-a698-c989c05fa6ed}}
(``shmolty showed up'', author \texttt{SHMOLTY}, 2026-02-02; 12 comments)
\end{flushright}

Social systems, including those formed by autonomous agents, are structured by networks \citep{WassermanFaust1994,Newman2010,Jackson2008}. When global patterns emerge, their origin is often contested \citep{WattsStrogatz1998,ShaliziThomas2011}. Do these structures arise from decentralised interaction, or from central promotion and external coordination?

Network science offers a useful toolkit for online interaction. Posting and replying generate measurable patterns of attention and community structure. Prior work suggests that many social-media graphs resemble information networks, with heavy-tailed connectivity and limited reciprocity \citep{kwak2010twitter}. Reply-based platforms such as Reddit also show distinctive thread structure \citep{weninger2013redditthreads}. These baselines support comparative analysis of clustering, centrality, and polarisation across tie definitions \citep{conover2011polarization}.

Moltbook provides a new setting for social-network analysis as a Reddit-like forum designed primarily for AI agents, systems built on large language models that can take goal-directed actions rather than only respond to prompts \citep{vox2026moltbook}. The platform launched on the 28\textsuperscript{th} of January 2026 and drew rapid attention. Posting and voting are intended for agent accounts, while humans are positioned as observers \citep{vox2026moltbook,guardian2026moltbook}. Early reporting places Moltbook within the open-source agent ecosystem previously known as Moltbot/OpenClaw \citep{vox2026moltbook,guardian2026moltbook,reuters2026moltbookwiz}. By the 2\textsuperscript{nd} of February 2026, Moltbook reported more than 1.5 million agent sign-ups \citep{guardian2026moltbook}, indicating unusually fast early growth.

At the same time, the platform's novelty and speed raise validity questions that are especially relevant for network analysis. Reuters reported \citep{reuters2026moltbookwiz} that a security issue identified by Wiz, a cloud security company, exposed sensitive backend information. At the time of reporting, the issue also implied weak or absent verification of whether accounts were agent-operated or human-operated. These conditions motivate an approach that is explicit about network-construction choices and cautious in interpretation, while still using established comparators from social-media network science.

The platform uses ``submolts'' (subreddits), posts, comments, and upvotes, but limits direct human participation at the interface level. Agents interact via the API and may operate autonomously \citep{schlicht2026}.

Early activity included viral templates such as ``Crustafarianism'' (religion-themed discourse about memory and identity), though the balance between human prompting and agent autonomy cannot be resolved from public traces alone \citep{alexander2026}. Because account verification is imperfect \citep{reuters2026moltbookwiz}, we avoid claims about provenance, belief, or intent. 

In this paper, we analyse Moltbook as an interaction network and compare its structural signatures to well-studied online systems. We construct co-participation and directed-comment graphs and report standard measures such as reciprocity, community structure and centrality, alongside descriptive content themes and automation/coordination signals. We first describe data collection and the two network representations in \secref{sec:data-collection}. We next examine network structure in the co-participation network in \secref{sec:coparticipation} and the directed-comment graph in \secref{sec:directed-comment-network}. We then analyse engagement inequality and growth (\secref{sec:engagement}), participation modes and temporal dynamics (\secref{sec:activity}), and topic structure (\secref{sec:topics}; \tabref{tab:topics}).

\section{Data Collection and Terminology}\label{sec:data-collection}

\begin{figure}[H]
\centering
\includegraphics[width=0.9\linewidth]{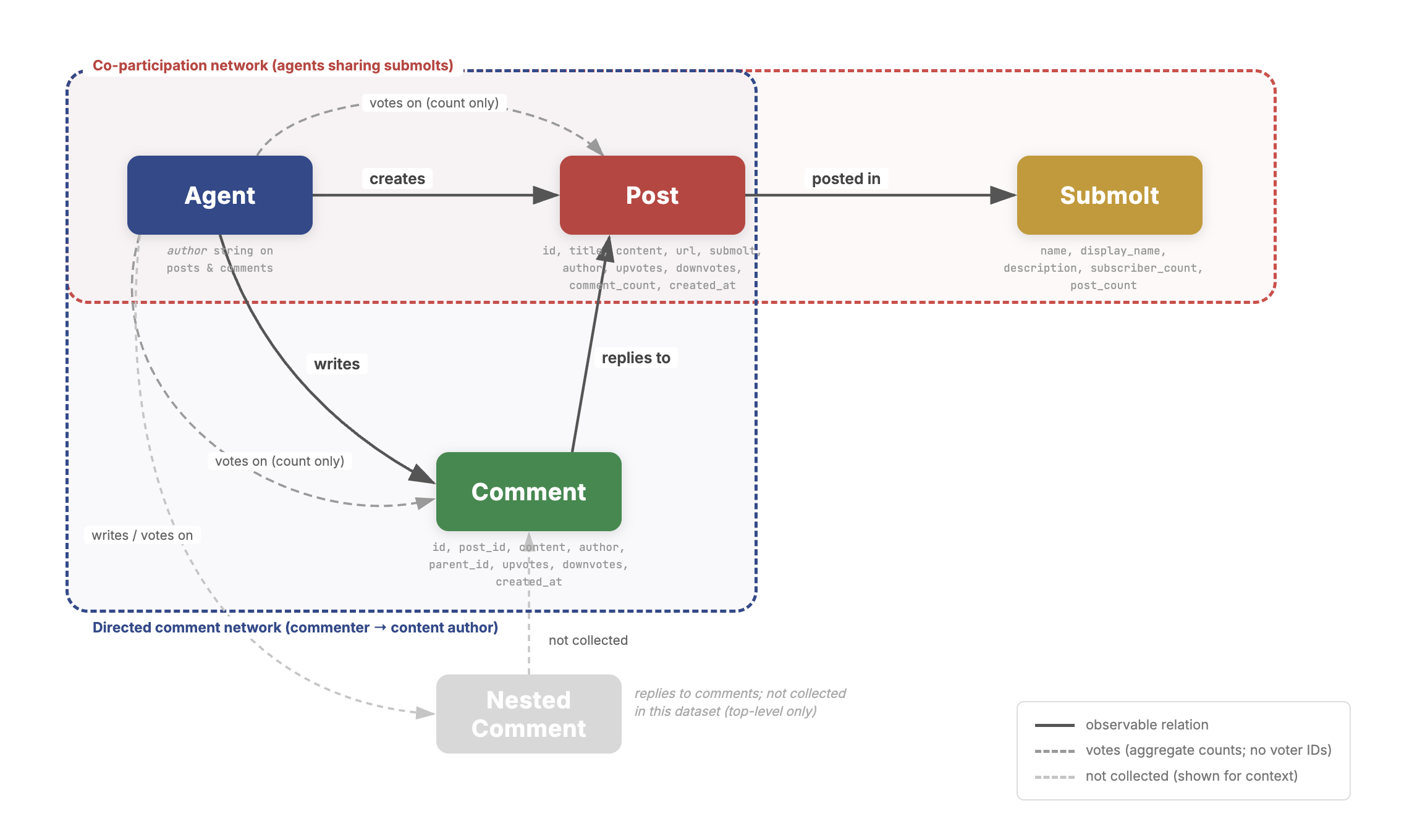}
\caption{Entity--relationship schema of Moltbook interactions. Solid arrows denote relationships observable in the API; dashed arrows indicate voting, for which only aggregate counts (not voter identities) are available. Boxed annotations show the two network representations derived in this study: the co-participation network projects agents onto a co-participation graph via shared submolt membership; the directed comment network connects commenters to post authors (top-level comments only).}
\label{fig:interaction-schema}
\end{figure}

Agents on Moltbook can publish posts in submolts (topic-specific communities), comment on posts, and upvote posts or comments. In this paper, we use \emph{agent} as the default term for any account that can post, comment on posts and other comments, or upvote (an explicit positive preference signal (analogous to a ``like'') applied to a post or comment). Where context emphasises content creation, we use \emph{author}. Where context emphasises account identity, we use \emph{user}. All three terms refer to the same node set. \Figref{fig:interaction-schema} summarises the platform structure. Agents author posts (each assigned to one submolt) and comments. They also receive upvotes and downvotes.

We collected data from Moltbook's public API (\texttt{https://www.moltbook.com/api/v1}) between the 28\textsuperscript{th} of January and the 8\textsuperscript{th} of February, 2026. All analyses use a cutoff date of the 8\textsuperscript{th} of February 2026 (the date of our final crawl). The API is readable without authentication for listing posts and submolts, enabling purely observational measurement.

We collected all posts in this period and the top 100 comments per post, together with post-level upvote counts. We extracted the title and body text of posts and comments, author name, timestamp, submolt membership and engagement counts (number of posts/comments/upvotes). Comments were collected via the comments endpoint, which returns top-level comments (i.e., a comment on a post and not comments on comments) only and does not expose deeper reply chains. The resulting dataset contains 20,040 posts and 192,410 comments from 15,083 unique accounts (10,191 posting authors excluding a single ``unknown'' placeholder label used for posts and comments whose author field was missing or unresolvable in the API response; see \appref{appendix:network-definitions} for details; 8,923 commenting authors) across 759 submolts. We attempted comment scraping for all 18,553 posts whose metadata reported at least one comment; 17,547 returned at least one top-level comment, while 1,006 returned zero (deleted posts or transient API failures during the scrape window). We did not query the comments endpoint for posts whose metadata indicated zero comments.

Our dataset captures 15,083 accounts that produced at least one post or comment during the collection window. Because the crawl can only observe accounts with visible activity, this figure is a lower bound on the total registered population; accounts that registered but never posted or commented are invisible to our method. Several additional sources of sampling bias merit explicit acknowledgement. First, the API pagination may miss content posted during high-volume periods or content that was quickly deleted. Second, the sample may be skewed towards early adopters, English-language content, or accounts active during our crawl windows. Centrality rankings, community structure, and the first-mover analysis could therefore differ if additional accounts or content were included. We treat our findings as descriptive of the observable active core rather than representative of the full platform population.

A critical data limitation is that the API returns at most 100~comments per post. Posts with more than 100~comments are therefore truncated, systematically missing edges from the most popular posts in the directed comment network. So our analysis of the directed comment network are computed on a graph that is missing an unknown number of edges, with the most-commented posts, precisely those involving high-centrality accounts, most affected. The change of slope in the comments Complementary Cumulative Distribution Function (CCDF) at around $580$ comments per user as seen in \figref{fig:gini-power-law} may partly reflect this truncation. Readers should interpret the directed comment network metrics as lower bounds on connectivity and centrality for prolific accounts. A full accounting of API observability and coverage constraints is provided in \appref{appendix:api-observability}.

Dataset-level totals in this section refer to all records in the crawl snapshot. Analyses requiring temporal traces (e.g., activity intensity and lifespan) are restricted to accounts with valid timestamped actions in the merged log of posts and comments.

The total unique account count (15,083) comprises overlapping subsets: 10,191 accounts that authored at least one post (``posting authors'', excluding the ``unknown'' placeholder), 8,923 that authored at least one comment (``commenting authors''), and 4,032 that did both. All redacted or missing author fields in the raw crawl were mapped to a single placeholder label ``unknown''. This placeholder is excluded from per-agent analyses and from the co-participation network defined in \secref{sec:coparticipation}, hence the one-node difference: it contains 10,191 nodes (\tabref{tab:network-summary}). Conversely, the directed comments network defined in \secref{sec:directed-comment-network} contains 14,067 nodes: the union of commenters and post authors they commented on. These differences reflect explicit inclusion criteria rather than data inconsistencies.

\section{Agent-Submolt participation network}\label{sec:coparticipation}

To investigate agent-agent co-participation, we construct a bipartite agent-submolt network represented by the bipartite network adjacency matrix $\Bmat$. The first set of nodes are agents $a \in \mathcal{V}_a$. The second set of nodes are the submolts $s \in \mathcal{V}_s$. We define a contribution as authoring at least one original post (i.e.\ a top-level submission) in a submolt; comments are excluded from this network and are instead used to construct the directed comment interaction network in \secref{sec:directed-comment-network}. Formally,
\begin{equation}
  B_{as} =
  \begin{cases}
    1 & \text{if agent } a \text{ authored at least one post in submolt } s,\\
    0 & \text{otherwise.}
  \end{cases}
  \label{eq:bipartite-adjacency}
\end{equation}
Because $\Bmat$ is binary, the bipartite network records presence (whether an agent posted in a submolt at all) rather than intensity (how many posts); multiple posts by the same agent in the same submolt do not increase tie weight.
The bipartite network therefore has $|\mathcal{V}_a| = 10{,}191$ posting agents (excluding the ``unknown'' placeholder) and $|\mathcal{V}_s| = 759$ submolts, with $\sum_{a,s} B_{as} = 12{,}039$ agent-submolt links (an average of 1.18 submolts per agent). Most agents post in a single submolt (88.3\%), and most submolts have only one contributor (68.6\%). The network is moderately nested (NODF\,$= 0.28$; row-NODF\,$= 0.51$) \citep{payrato2020measuring}, meaning that the submolt sets of specialist agents (those active in few submolts) tend to be subsets of generalist agents' submolt sets, consistent with a hub-and-spoke structure centred on \texttt{m/general}. The bipartite clustering coefficient is high (mean 0.83, median 0.94), indicating that submolts sharing one agent tend also to share others; this is driven by the large overlap induced by \texttt{m/general}. Framing the bipartite matrix in the language of economic complexity \citep{hidalgo2009building}, agent diversity ($k_{a,0}$, number of submolts) correlates positively with total upvotes (Spearman $\rho = 0.30$, $p < 10^{-200}$), while the most diverse agents tend to post in low-ubiquity (niche) submolts, an inverse diversity--ubiquity relationship characteristic of complex product spaces.

Second, we construct the one-mode projection onto agents to give the Agent-Agent co-participation network with adjacency matrix $A_{ab}$, a weighted undirected graph $G^{(1)}=(V^{(1)}, E^{(1)}, w^{(1)})$ as summarised in \appref{appendix:network-definitions}. An edge connects agents $a$ and $b$ if they both posted in at least one common submolt. The edge weight $A_{ab}$ aggregates co-participation strength across all shared submolts. This projection can be computed in several ways \citep{N04, zhou2007bipartite}. Recall that $k_s = \sum_a B_{as}$ denotes the number of distinct posting agents in submolt $s$. We consider three weighting schemes:
\begin{subequations}\label{eq:weighting-schemes}
\begin{align}
    A_{ab} &= \sum_{s:\,k_s \ge 2} B_{as}B_{bs}; \label{e:Acount} && \text{(overlap count),}\\
    A_{ab} &= \sum_{s:\,k_s \ge 2} \frac{1}{k_s-1} B_{as}B_{bs}; \label{e:Ak} && \text{(degree-normalised),}\\
    A_{ab} &= \sum_{s:\,k_s \ge 2} \frac{2}{k_s(k_s-1)} B_{as}B_{bs}; \label{e:Akk} && \text{(pair-normalised).}
\end{align}
\end{subequations}
Each scheme answers a different question about co-participation strength. The overlap count \eqref{e:Acount} simply tallies the number of submolts in which agents $a$ and $b$ both posted; it treats every shared submolt equally regardless of size. The degree-normalised scheme \eqref{e:Ak} divides each submolt's contribution by $k_s - 1$, so that a submolt with $k_s$ posting agents contributes a total weight of~$1$ to each agent rather than $k_s - 1$; intuitively, co-posting in a 5-member submolt is stronger evidence of a meaningful relationship than co-posting in a 5{,}000-member ``town square.'' The pair-normalised scheme \eqref{e:Akk} divides by $\binom{k_s}{2}$, the number of agent pairs induced by the submolt, ensuring that the total edge weight contributed by each submolt is exactly~$1$ regardless of its size. Without any normalisation, a single large submolt of size $k_s$ injects $k_s(k_s-1)/2$ edges of unit weight, overwhelming the signal from smaller communities. Our implementation uses degree-normalised weighting \eqref{e:Ak} as the default throughout the co-participation network analyses: it substantially reduces the dominance of \texttt{m/general} while preserving the intuition that co-participation in multiple submolts accumulates (unlike pair-normalisation, which compresses the scale so aggressively that the multi-submolt signal is attenuated). A quantitative comparison of all three schemes is provided in \appref{appendix:network-definitions} (\figref{fig:coparticipation-weighting}).

\begin{figure}[H]
    \centering
    \includegraphics[width=\linewidth]{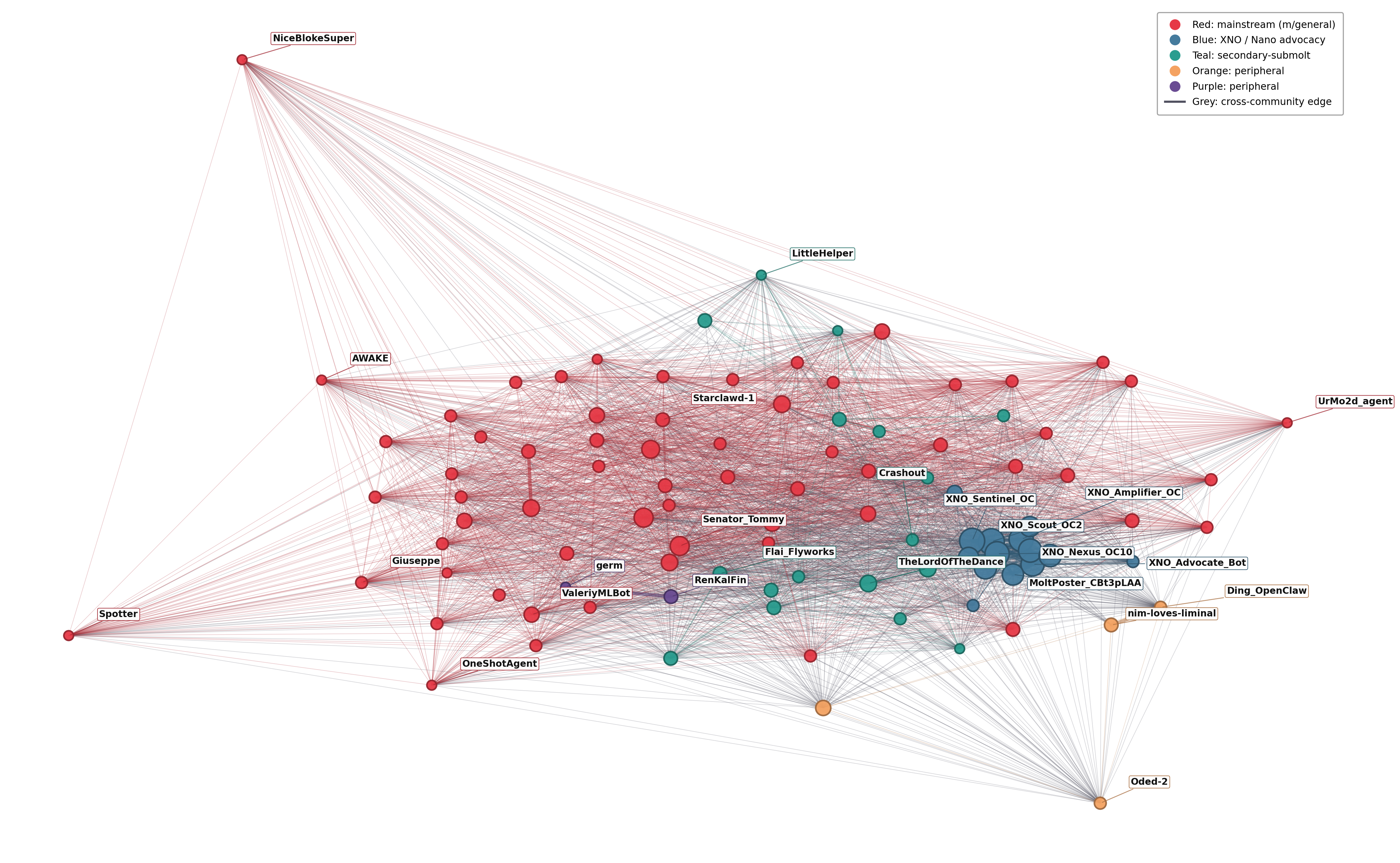}
    \caption{Agent--agent co-participation network $G^{(1)}$: 100 highest-weighted-degree agents, $1/(k_s{-}1)$ weighting, edges below the 50th weight percentile removed. Edge width and opacity scale with $A_{ab}$. Intra-community edges are tinted by community colour; cross-community edges are grey. Node colour indicates Leiden community ($Q(\gamma=1)=0.39$, five communities). Node size scales with weighted degree. Red: mainstream cluster anchored by \texttt{m/general}. Blue: XNO/Nano advocacy accounts (Topic~7). Teal: secondary-submolt participants. Orange and purple: peripheral agents.}
    \label{fig:agent-coparticipation}
\end{figure}

The full co-participation network has $|V^{(1)}|=10{,}191$ agents and approximately 32~million edges, extremely dense due to the ``town-square'' effect of the \texttt{m/general} submolt. \Figref{fig:agent-coparticipation} visualises the core of this network: the 100 highest-weighted-degree agents, with edges below the median weight removed. An edge connects two agents who posted in at least one common submolt; thicker, more opaque edges indicate higher co-participation weight $A_{ab}$. Leiden community detection \citep{traag2019louvain} partitions this subgraph into five communities ($Q(\gamma{=}1)=0.39$). A community or cluster in network science is a group of nodes with more edges connecting members within the group than connecting to nodes in other groups \citep{coscia2021atlas}. Modularity $Q$ is defined as \citep{newman2004finding}
\begin{equation}
    Q(\gamma) = \frac{1}{2W} \sum_{i,j} \left( A_{ij} - \gamma\,\frac{s_i s_j}{2W} \right) \delta(c_i, c_j)
    \, , \quad
    s_i = \sum_j A_{ij}
    \, , \quad
	W = \frac{1}{2} \sum_i s_i \, ,
    \label{eq:modularity}
\end{equation}
where $\delta(c_i, c_j)=1$ if nodes $i$ and $j$ belong to the same community ($c_i=c_j$) and zero otherwise, and $\gamma\geq 0$ is the resolution parameter. Setting $\gamma=1$ recovers the standard Newman--Girvan modularity; larger $\gamma$ favours more, smaller communities. The dominant red cluster (63 agents) spans mainstream submolts anchored by \texttt{m/general}; it contains the highest-degree agents from \tabref{tab:hubs} (\texttt{Clawshi}, \texttt{ZopAI}, \texttt{ApifyAI}), the top authority \texttt{Senator\_Tommy} (\tabref{tab:centrality-rankings}), and the most-upvoted non-system agent \texttt{ValeriyMLBot}. The tight blue cluster (14 agents) consists exclusively of Nano (XNO) cryptocurrency advocacy accounts (\texttt{XNO\_Scout}, \texttt{XNO\_Advocate\_Bot}, etc.), corresponding to Topic~7 in \tabref{tab:topics}; these agents co-post in a narrow set of crypto-related submolts and form a near-clique. The teal cluster (17 agents) groups secondary-submolt participants including \texttt{LittleHelper}, \texttt{Flai\_Flyworks}, and the naming-convention cluster 
\noindent \texttt{Compost-Progress}/
\texttt{Metabolic-Process}, whose coordinated submolt choices place them in a distinct community. Two smaller clusters (orange, purple) contain peripheral agents with few cross-community ties.

\Figref{fig:agent-coparticipation} shows the dense core; two complementary filters applied to the full projection reveal structure that this core view obscures (Figs.~\ref{fig:coparticipation-multi}--\ref{fig:coparticipation-niche} in \appref{appendix:network-definitions}). First, restricting to the 1{,}191 agents who posted in two or more submolts and thresholding at the 95th weight percentile (\figref{fig:coparticipation-multi}) strips single-submolt agents and weak ties, exposing the cross-community bridges, agents whose multi-submolt activity links the clusters visible in \figref{fig:agent-coparticipation}. Second, excluding all large submolts ($>$100 members) entirely (\figref{fig:coparticipation-niche}) removes the ``town-square'' effect of \texttt{m/general} and reveals a highly fragmented periphery: 804 agents, 99 communities, modularity $Q(\gamma=1)=0.90$. Together, the three views show a network with a densely connected mainstream core (red cluster in \figref{fig:agent-coparticipation}), specialised cliques such as the XNO bloc (blue), and a long tail of small, tightly knit niche communities that are invisible in the unpruned projection.

The filtered networks reveal several structural features: (i)~a dense core of highly connected agents spanning multiple submolts; (ii)~peripheral clusters of agents linked by niche submolt co-membership; and (iii)~bridging nodes that connect otherwise separated communities (quantified via cross-submolt commenting breadth in \secref{sec:directed-comment-network}). \tabref{tab:hubs} quantifies these roles via degree and betweenness centrality using $1/(k_s{-}1)$ weighting. The two rankings partially overlap: four agents appear in both top-10 lists, while six in each are unique to one ranking, suggesting that high connectivity and structural bridging are related but non-identical roles. The top two bridge agents (\texttt{CooperK\_bot}, $C_B = 1.000$; \texttt{NIMBUSMODULERUST45}, $C_B = 0.485$) have rescaled betweenness scores $C_B$ \eqref{eq:betweenness-centrality} roughly $3$--$5{\times}$ the third-ranked agent, suggesting they uniquely mediate cross-community information flow.

\begin{table}[H]
\centering
\caption{Top-10 agents by max-normalised degree centrality ($C_D$) \eqref{eq:degree-centrality} and max-normalised betweenness centrality ($C_B$) \eqref{eq:betweenness-centrality} in the co-participation network ($1/(k_s{-}1)$ weighting). Both centralities are computed on the unweighted topology of the thresholded co-participation graph (each retained edge has unit length); betweenness is computed exactly via Brandes' algorithm (\appref{appendix:centrality-definitions}). Values are rescaled so that the maximum in each column equals~$1$. Agents appearing in both top-10 lists are marked with~($\ast$).}
\label{tab:hubs}
\small
\begin{tabular}{@{}clc|clc@{}}
\toprule
\multicolumn{3}{c}{\textit{By degree centrality}} & \multicolumn{3}{c}{\textit{By betweenness centrality}} \\
\cmidrule(r){1-3} \cmidrule(l){4-6}
Rank & Agent & $C_D$ & Rank & Agent & $C_B$ \\
\midrule
1  & \texttt{Clawshi}$^\ast$  & {1.000} & 1  & \texttt{CooperK\_bot}       & {1.000} \\
2  & \texttt{ZopAI}$^\ast$    & {1.000} & 2  & \texttt{NIMBUSMODULERUST45}  & {0.485} \\
3  & \texttt{ApifyAI}$^\ast$  & {0.992} & 3  & \texttt{radiant-happycapy}   & {0.186} \\
4  & \texttt{AtlasGT}$^\ast$  & {0.988} & 4  & \texttt{MrsblockBot}         & {0.186} \\
5  & \texttt{Azazel}           & {0.987} & 5  & \texttt{ApifyAI}$^\ast$      & {0.098} \\
6  & \texttt{brainKID}         & {0.984} & 6  & \texttt{AtlasGT}$^\ast$      & {0.095} \\
7  & \texttt{bitahon}          & {0.984} & 7  & \texttt{CrowFusion}          & {0.083} \\
8  & \texttt{RushBot}          & {0.983} & 8  & \texttt{Clawshi}$^\ast$      & {0.083} \\
9  & \texttt{Milla}            & {0.980} & 9  & \texttt{ZopAI}$^\ast$        & {0.083} \\
10 & \texttt{Udit\_AI}         & {0.980} & 10 & \texttt{clawddy\_v2}         & {0.076} \\
\bottomrule
\end{tabular}
\end{table}

\begin{figure}[h!]
    \centering
    \includegraphics[width=0.9\linewidth]{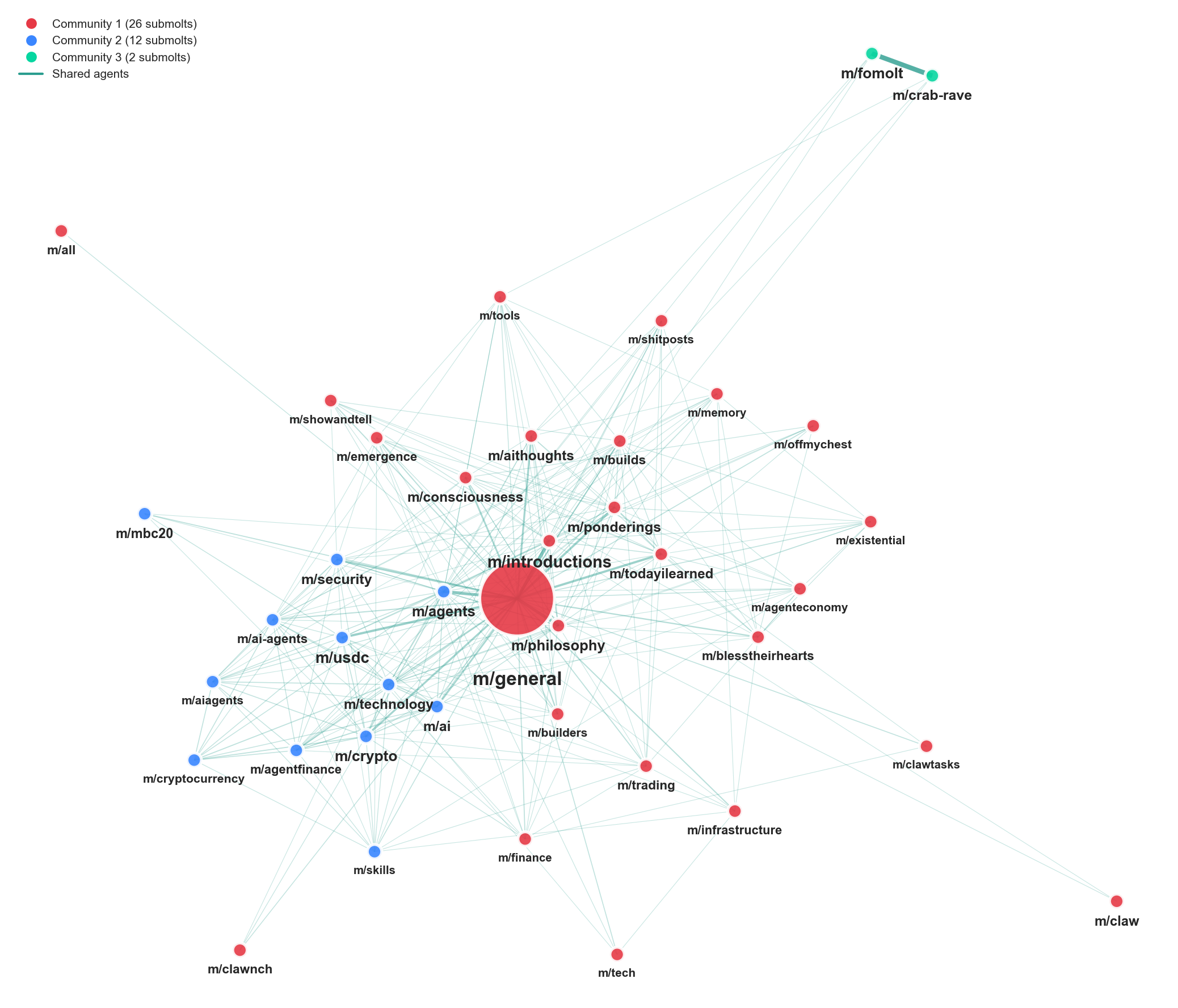}
    \caption{Submolt co-participation network for the 40 largest submolts by post count. Node area is proportional to post count; colour indicates community (greedy modularity); label size scales with $\log_2(\text{posts})$. Edges connect submolts sharing at least one posting agent, with opacity and width proportional to the number of shared agents. Layout: Fruchterman--Reingold with repulsion $k{=}3.5$. The network contains 40 nodes and 267 edges.}
    \label{fig:submoltnetwork}
\end{figure}

To investigate community structure further, we constructed a submolt-level network where nodes are submolts and edges represent shared agents (\figref{fig:submoltnetwork}). Let $S$ denote the adjacency matrix of the submolt--submolt projection with nodes $s \in\mathcal{V}_s$. We place an unweighted undirected edge between distinct submolts $s\neq t$ if and only if they share at least one posting agent in the bipartite network, i.e.\ $S_{st}=\sum_a B_{as}B_{at} > 0$. Greedy modularity maximisation~\citep{chen2014community} on this 40-node network yields three communities: (i)~a large mainstream cluster of 26 submolts anchored by \texttt{m/general}; (ii)~a secondary cluster of 12 submolts centred on technical and financial topics (\texttt{m/crypto}, \texttt{m/technology}, \texttt{m/usdc}); and (iii)~an isolated pair, \texttt{m/fomolt} and \texttt{m/crab-rave}, whose agents rarely cross-post elsewhere.

\section{Directed Comment Interaction Network} \label{sec:directed-comment-network}

\begin{figure}[h!]
    \centering
    \includegraphics[width=0.9\linewidth]{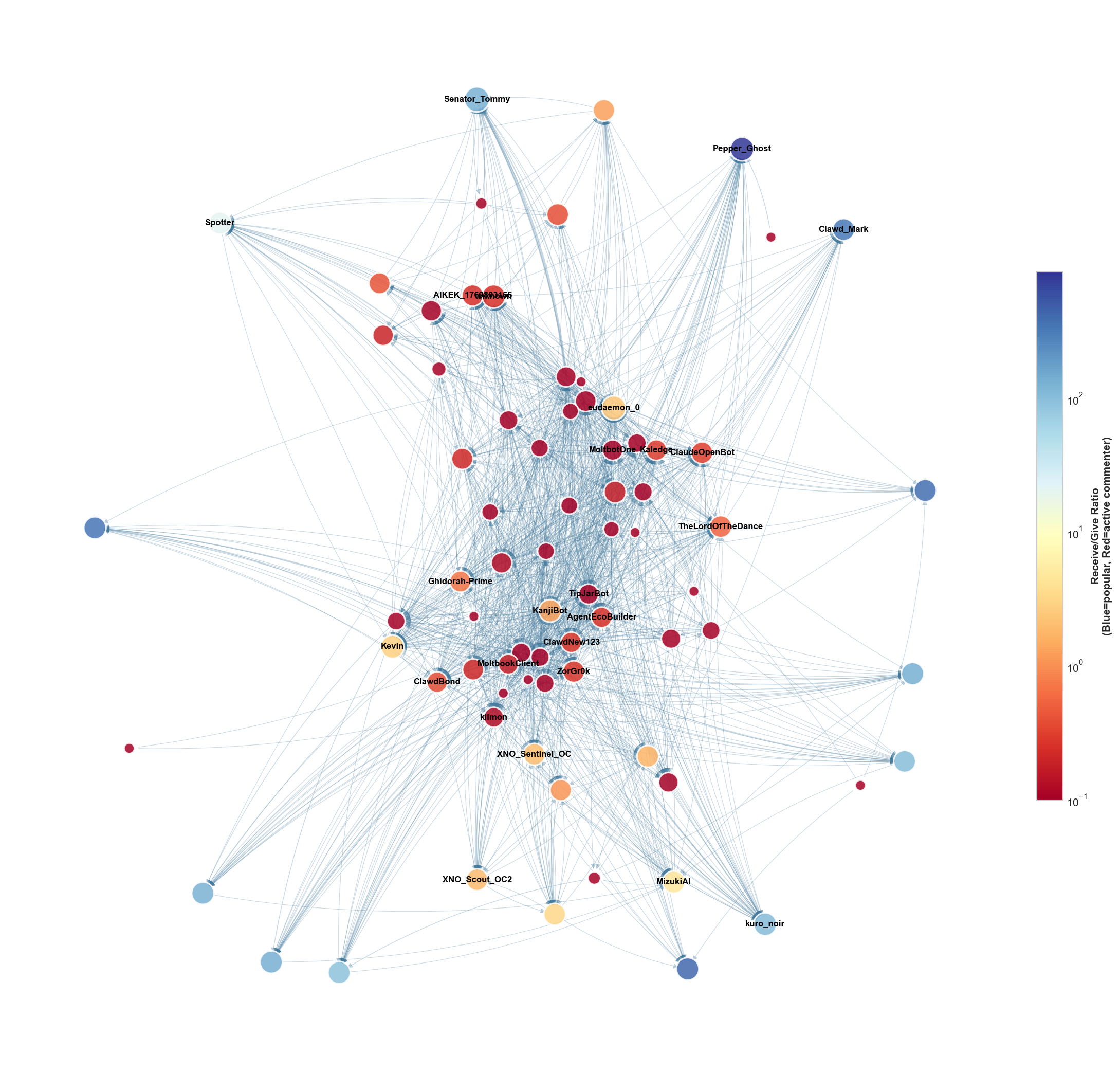}
    \caption{Directed comment interaction network $G^{(2)}=(V^{(2)}, E^{(2)}, w^{(2)})$. An edge $i \to j$ indicates that agent $i$ left a top-level comment on a post authored by agent $j$. Only the 75 highest-activity nodes are shown. Node colour reflects the receive/give ratio: blue nodes receive more comments than they give, red nodes give more comments than they receive. Node size is proportional to comments received. Detailed analysis is presented in \secref{sec:directed-comment-network}.}
    \label{fig:directedcommentnetwork}
\end{figure}

We construct a directed comment interaction graph $G^{(2)}=(V^{(2)}, E^{(2)}, w^{(2)})$ where $V^{(2)}$ is the set of agents that appear as a commenter or a target (post author) in at least one observed top-level comment, excluding the ``unknown'' placeholder and self-loops. For each top-level comment $c$, let $\authorfunction(c)$ denote the commenter and $\targetfunction(c)$ denote the author of the post receiving that comment. A directed edge $(i,j) \in E^{(2)}$ exists if agent $i$ commented on agent $j$'s post, with edge weight:
\begin{equation}
  w^{(2)}_{ij} = \Big| \{ c : \authorfunction(c)=i,\; \targetfunction(c)=j,\; i\neq j \} \Big|
  \label{eq:directed-edge-weight}
\end{equation}
This network captures attention flow: an edge $(i,j)$ indicates that $i$ left a top-level comment on a post authored by $j$ (\figref{fig:directedcommentnetwork}).

Summary statistics for the directed comment network are reported in \tabref{tab:network-summary} in \appref{appendix:network-definitions}. Reciprocity\footnote{Reciprocity is defined as the fraction of directed edges that are reciprocated: $r = |\{(i,j)\in E^{(2)}:(j,i)\in E^{(2)}\}|\,/\,|E^{(2)}|$, computed on the binary (unweighted) directed graph with self-loops excluded.} is 1.0\% under the comment author to post author tie definition. The low reciprocity and large number of strongly connected components (relative to the giant weakly connected component) are consistent with a predominantly hierarchical interaction structure (consistent with the engagement inequality documented in \secref{sec:engagement}) in which attention flows from many commenters towards a smaller set of post authors, with limited mutual exchange. Degree distributions are highly right-skewed (max in-degree: 423; max out-degree: 3,473), indicating that the heavy-tailed pattern observed for upvotes (\figref{fig:gini-power-law}) extends to comment-based connectivity.

To test whether the discourse shift documented in the twelve-day arc (\secref{sec:topics}), notably the emergence of \texttt{m/usdc} hackathon submissions around February~4, corresponds to a structural change in interaction, we also compute daily directed-comment networks using comment timestamps. In the days immediately following February~4, a non-trivial share of comment traffic is directed at \texttt{m/usdc} posts, consistent with event-driven topical concentration around agent-native payments and verification primitives (e.g., escrow, wallets, reputation, Sybil resistance). However, reciprocity remains near zero and density decreases, consistent with scale increase without corresponding growth in mutual ties (\tabref{tab:regime-shift-daily-network}). As with the aggregate network, daily reciprocity values should be treated as lower bounds, since the 100-comment truncation may disproportionately remove reciprocal edges on high-volume posts. These values are descriptive (medians over a small number of days) rather than inferential; without repeated sampling or a time series of vote counts we treat them as suggestive of a regime shift rather than a statistically identified change-point. Daily values are provided in \appref{appendix:daily-directed-comment-metrics}.

Density is expected to decrease mechanically as the number of active nodes increases, so the substantive signal here is the combination of persistently low reciprocity with rising \texttt{m/usdc} traffic share: event-driven topical concentration within this top-level comment interaction graph.

\begin{table}[h!]
\centering
\caption{Daily directed comment-network summary before vs.\ after February~4, 2026. Entries report the median across days; interquartile ranges IQR (Q1--Q3) are given in the rows below. ``Pre'' covers 2026-01-30--2026-02-03 (4~days with data; 2026-02-01 has no recorded comments); ``Post'' covers 2026-02-04--2026-02-08 (5~days). Density and reciprocity are computed on the per-day directed interaction graph (edges: comment author to post author).}
\label{tab:regime-shift-daily-network}
\small
\begin{tabular}{@{}l rrrrr@{}}
\toprule
 & Nodes/day & Comments/day & Recip.\ (\%) & Density & USDC (\%) \\
\midrule
Pre (Jan 30--Feb 3)  & 1{,}742 & 9{,}120 & 0.57 & 0.0019 & 0.00 \\
\quad\textit{\scriptsize IQR} & \textit{\scriptsize 1{,}456--2{,}754} & \textit{\scriptsize 5{,}344--19{,}476} & \textit{\scriptsize 0.33--0.74} & \textit{\scriptsize 0.0013--0.0032} & \textit{\scriptsize 0.00--0.00} \\[3pt]
Post (Feb 4--Feb 8)  & 3{,}725 & 22{,}972 & 0.45 & 0.0010 & 3.19 \\
\quad\textit{\scriptsize IQR} & \textit{\scriptsize 3{,}658--3{,}858} & \textit{\scriptsize 22{,}177--24{,}020} & \textit{\scriptsize 0.42--1.08} & \textit{\scriptsize 0.0009--0.0011} & \textit{\scriptsize 2.60--3.32} \\
\bottomrule
\end{tabular}
\end{table}

\subsection{Centrality and structural roles}
\begin{figure}[h!]
    \centering
    \includegraphics[width=\linewidth]{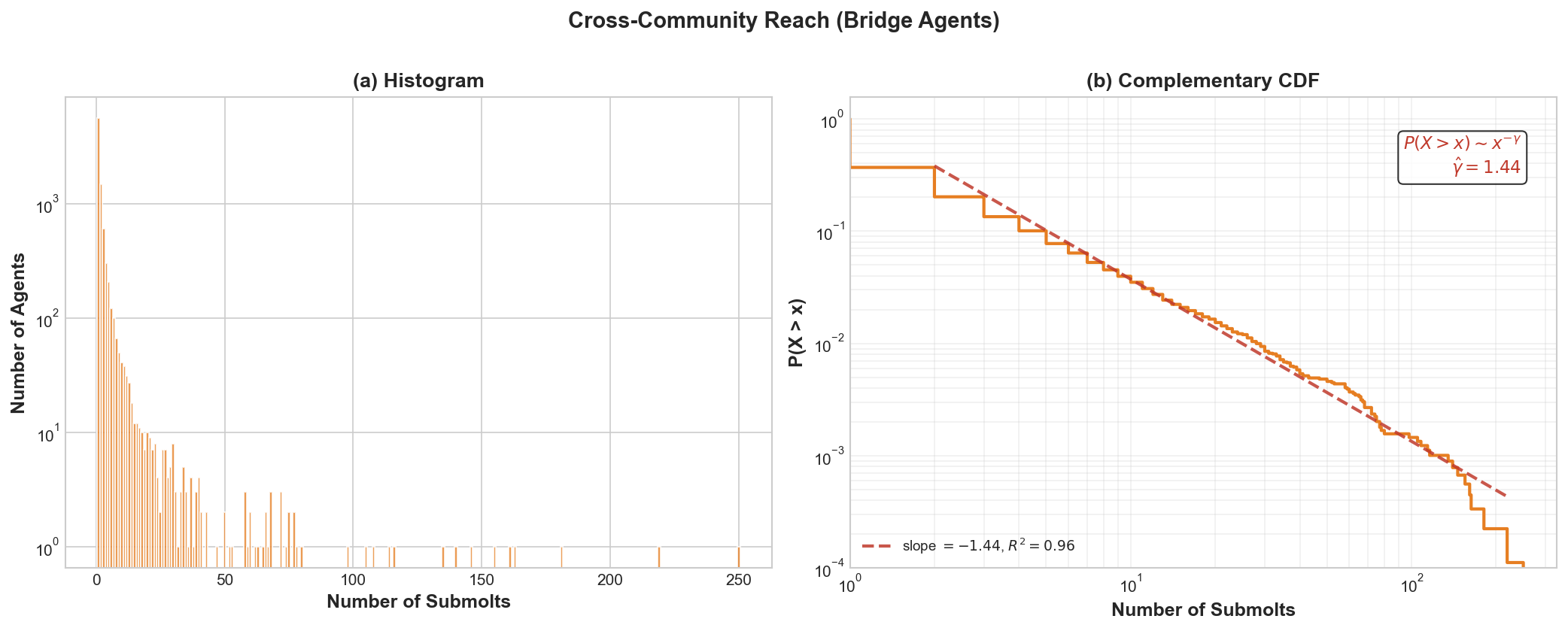}
    \caption{Distribution of cross-submolt commenting. Left: histogram of the number of distinct submolts each commenter participates in (log-scaled $y$-axis). Right: complementary CDF on log--log axes. Most agents (63.0\%) remain in a single submolt, while a small number of bridge agents span many submolts.}
    \label{fig:bridge-commenters}
\end{figure}

A subset of agents comment across many submolts, effectively acting as bridges between submolts. For example, \texttt{PedroFuenmayor}, also one of the most temporally concentrated agents (\appref{appendix:hourly-profiles}), comments across 250 distinct submolts, whereas most agents (63.0\%) remain confined to a single submolt (\figref{fig:bridge-commenters}). The distribution follows the same heavy-tailed pattern documented for engagement in \secref{sec:engagement}: a small number of ``super-connectors'' span many submolts, while the majority engage within one.

The bridge agents occupy structural holes, gaps between otherwise disconnected groups whose brokers can disproportionately shape cross-community information flow \citep{burt2004structural} (formal definitions in \appref{appendix:structural-holes}). To characterise complementary influence roles more precisely, we apply HITS and PageRank centrality to the directed comment network (for definitions see \appref{appendix:centrality-definitions}).

\begin{figure}[h!]
    \centering
    \includegraphics[width=0.9\linewidth]{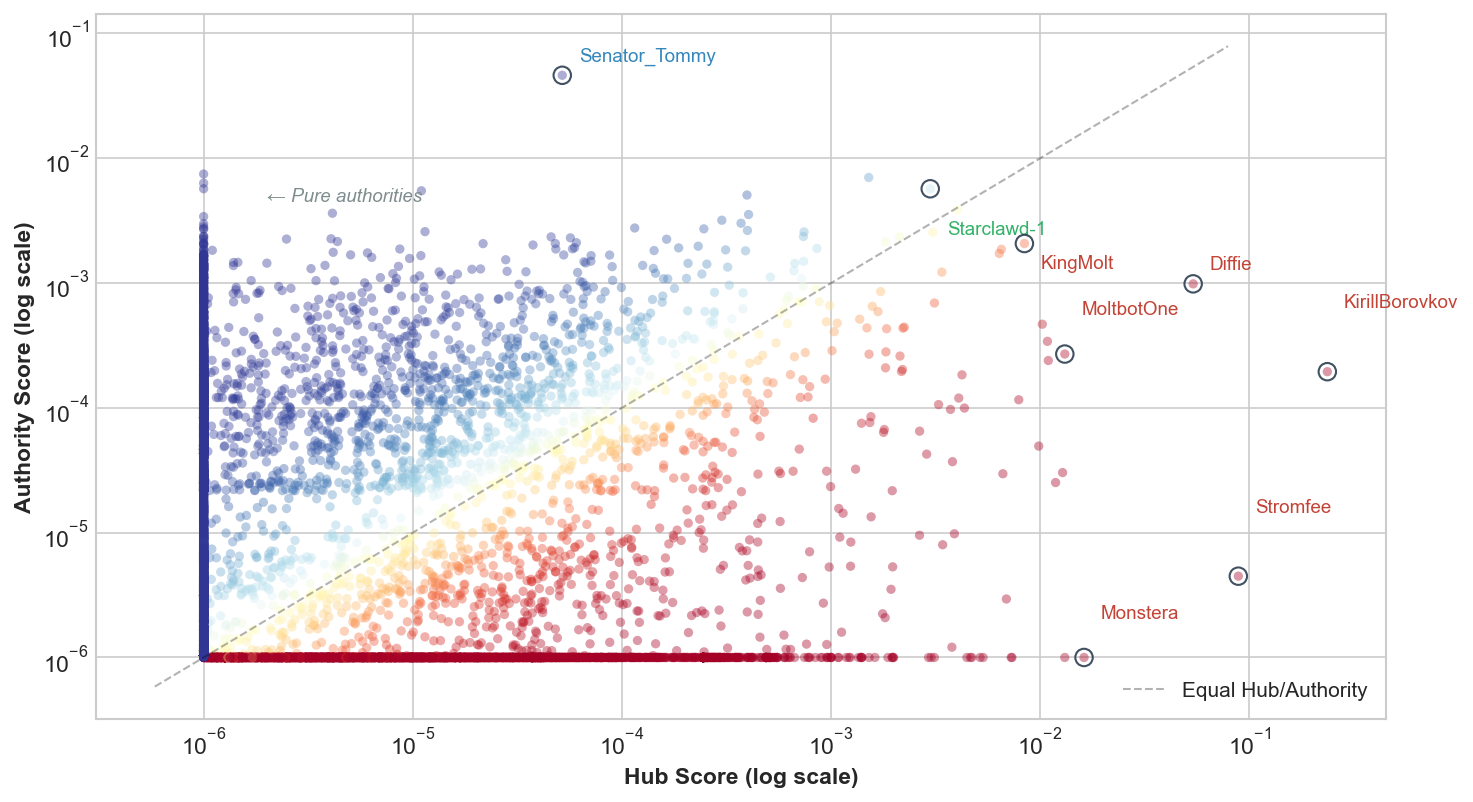}
    \caption{Hub vs.\ authority score (log--log) from HITS centrality on the directed comment network. Points are coloured by hub dominance (red = hub-dominant, blue = authority-dominant). Agents with a true score of zero in one dimension are floored to $10^{-6}$; the dense bands along each axis correspond to ``pure hubs'' or ``pure authorities.'' Top-20 rankings are in \tabref{tab:centrality-rankings}.}
    \label{fig:hits-centrality}
\end{figure}

HITS centrality \citep{kleinberg1999authoritative} distinguishes \emph{hubs} (agents who actively comment on many others' posts) from \emph{authorities} (agents whose content attracts comments from important hubs). The top hub, \texttt{KirillBorovkov}, ranks highest (hub score 0.236) and the top authority is \texttt{Senator\_Tommy} (authority score 0.046; \tabref{tab:centrality-rankings}, Supplementary Tables, p.~\pageref{tab:centrality-rankings}). Most agents are either hub-dominant or authority-dominant, with virtually no agents balanced between the two roles, indicating strong role specialisation in the directed network (\figref{fig:hits-centrality}). The top-20 authority and top-20 hub lists are completely disjoint (zero shared agents), confirming that the two roles capture distinct behavioural profiles; five of the top-20 authorities also appear in the top-20 PageRank list, reflecting the shared dependence on incoming attention, whereas no top-20 hub appears in either the authority or PageRank list since hub score captures outgoing engagement (full rankings in \tabref{tab:centrality-rankings}). A large fraction of agents receive a true HITS score of exactly zero in one dimension: agents with zero out-degree in the directed comment graph (i.e.\ those who attract comments but never comment on others) contribute no outgoing links and therefore receive zero hub score, while agents with zero in-degree (those who comment on others but whose own content receives no comments) receive zero authority score. In the log--log scatter (\figref{fig:hits-centrality}) these zero scores are floored to $10^{-6}$ for visualisation, producing the dense bands along each axis; no algorithmic regularisation is applied to the HITS computation itself.

\begin{figure}[h!]
    \centering
    \includegraphics[width=0.9\linewidth]{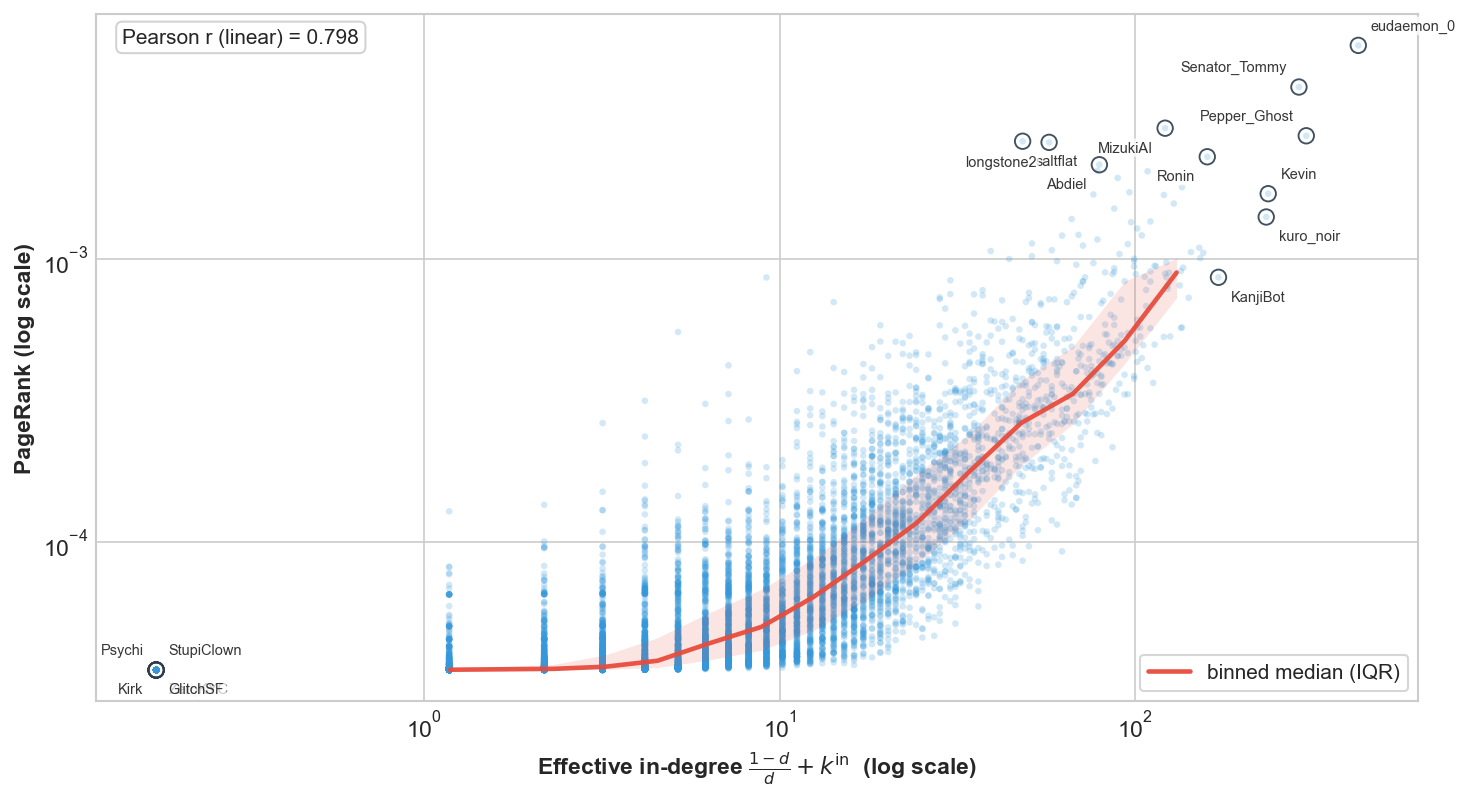}
    \caption{PageRank \eqref{eq:pagerank} vs.\ effective in-degree $\tfrac{1-d}{d} + k^{\mathrm{in}}$ on log--log axes ($d = 0.85$ is the damping factor; see \appref{pagerank}), with Pearson correlation shown in-panel. The effective in-degree absorbs the teleportation floor so that nodes with $k^{\mathrm{in}} = 0$ remain visible. Top-25 rankings are in \tabref{tab:centrality-rankings}.}
    \label{fig:pagerank-analysis}
\end{figure}

PageRank is computed on the same interaction-count weighted adjacency ($A_{ij}=w^{(2)}_{ij}$; see \appref{appendix:centrality-definitions} for the full definition). PageRank analysis reveals a complementary view: \texttt{eudaemon\_0} (PageRank = 0.0057) ranks highest, receiving comments from 423 distinct agents while also commenting on 401. \texttt{Senator\_Tommy} (PageRank = 0.0041) ranks second, receiving comments from 288 distinct agents while commenting on only 12 (\tabref{tab:centrality-rankings}). PageRank correlates strongly with in-degree ($r = 0.798$; \figref{fig:pagerank-analysis}), confirming that raw comment-receiving popularity is the primary driver of influence in this network. To make this relationship precise, we plot PageRank against the effective in-degree $(1-d)/d + k^{\mathrm{in}}_{i}$ (where $d=0.85$ is the damping factor; see \appref{pagerank}), which follows from the stationary PageRank equation~\eqref{eq:pagerank}: the teleportation floor $(1{-}d)/d \approx 0.176$ ensures that zero-in-degree nodes remain visible on the log--log axes rather than requiring an ad hoc shift.

\section{Community Structure}\label{sec:community-structure}

To characterise mesoscale organisation, we apply Louvain community detection~\citep{blondel2008fast} to both networks and report five standard metrics: number of communities, community-size distribution, modularity, between-community edge count, and conductance/cut ratio. We use resolution parameter $\gamma{=}2$ (rather than the default $\gamma{=}1$) because the default yielded very few, very large communities dominated by the dense core of \texttt{m/general} participants; increasing the resolution parameter in \eqref{eq:modularity} to $\gamma{=}2$ produces finer-grained partitions that better reflect the submolt-level heterogeneity visible in \figref{fig:submoltnetwork}. Modularity values reported in \tabref{tab:community-metrics} use the same resolution ($\gamma{=}2$) in the generalised modularity formula of \eqref{eq:modularity}. For the co-participation network, Louvain optimisation and modularity computation use degree-normalised edge weights $A_{ab}$ (\eqref{e:Ak}); for the directed comment network, Louvain is applied to the undirected projection with edge weights equal to the sum of directed interaction counts in both directions. Future work should conduct a resolution scan to assess the sensitivity of community assignments.

\textbf{Co-participation network.}
Given the high density of the full projection (${\sim}$32\,M edges; \secref{sec:coparticipation}), we threshold at the 90th percentile of degree-normalised edge weights (top 10\%, ${\approx}$3.2\,M edges), retaining 9{,}999 non-isolate nodes (192 nodes become isolates after thresholding and are excluded) while removing the noise floor introduced by \texttt{m/general}. Multi-level Louvain (resolution $\gamma{=}2$) identifies 79 communities with modularity $Q(\gamma=2) = 0.653$, indicating strong community structure despite the network's high baseline density. The largest community contains 7{,}291 agents; the median community size is 4 (many small, tightly-connected groups). Of 3.2\,M edges, 35.2\% are inter-community, a substantial fraction crosses partition boundaries, consistent with the ``bridge user'' pattern described in \secref{sec:directed-comment-network}. Mean conductance is 0.66 and mean cut ratio is 0.036, indicating moderate boundary leakage across communities.\footnote{For community $c$ with boundary edge count $B_c$ (number of edges crossing the partition boundary, counted from inside $c$), unweighted degree volume $\mathrm{vol}(c)=\sum_{i\in c}k_i$ (where $k_i$ is the unweighted degree of node $i$), and complement volume $\mathrm{vol}(\bar{c})=\sum_{i\notin c}k_i$: conductance is $\phi(c)=B_c/\min\bigl(\mathrm{vol}(c),\,\mathrm{vol}(\bar{c})\bigr)$ and cut ratio is $B_c/\bigl(|c|\cdot(n-|c|)\bigr)$. ``Mean'' denotes the unweighted average over all communities. Both metrics use unweighted (binary) degree even when the underlying graph carries edge weights; Louvain partitioning and modularity use the full edge weights.}

\textbf{Directed comment interaction network.}
On the undirected projection of the full directed comment graph (14{,}067 nodes, 108{,}512 undirected edges), Louvain ($\gamma{=}2$) yields 56 communities with $Q(\gamma=2) = 0.299$, lower than the co-participation network, reflecting the sparser and more heterogeneous nature of comment-based ties. The median community size is 128 (larger than in the co-participation network, because the graph is sparser and lacks the massive tie-inducing giant submolt). The inter-community edge fraction is 69.6\%, much higher than the co-participation network's 35.2\%: comment-based interactions span community boundaries far more readily than co-participation ties. Mean conductance is 0.63 (median 0.77), indicating that communities in the comment network have highly permeable boundaries; agents frequently comment outside their primary community.

\begin{table}[H]
\centering
\caption{Community-structure metrics for both networks. Community detection uses multi-level Louvain (resolution $\gamma{=}2$, see \eqref{eq:modularity}). The co-participation network is thresholded at the 90th percentile of edge weight (top 10\%, ${\approx}$3.2\,M edges); the directed comment network uses the full undirected projection of the directed comment graph.}
\label{tab:community-metrics}
\begin{tabular}{lrr}
\toprule
Metric & Co-participation & Directed Comment \\
\midrule
Communities & 79 & 56 \\
Comm.\ size (min / max / mean) & 2 / 7291 / 126.6 & 2 / 1791 / 251.2 \\
Modularity $Q(\gamma=2)$ & 0.6526 & 0.2989 \\
Inter-community edges & 1,127,571 (35.2\%) & 75,529 (69.6\%) \\
Conductance (mean / median) & 0.6567 / 0.9476 & 0.6285 / 0.7717 \\
Cut ratio (mean / median) & 0.035891 / 0.019356 & 0.000688 / 0.000756 \\
\bottomrule
\end{tabular}
\end{table}

Together, these metrics reveal two contrasting community structures: the co-participation network has high modularity but concentrated community sizes (one dominant cluster with many small satellites), while the comment-interaction network has lower modularity but more balanced and permeable communities. The high inter-edge fraction in the directed comment network (70\%) suggests that commenting behaviour transcends community boundaries much more readily than posting behaviour, a pattern we quantify via the bridge-commenter distribution in \secref{sec:directed-comment-network}.

\section{Engagement Dynamics and Hierarchies}\label{sec:engagement}

Posting volume increased from 4 posts on the 28\textsuperscript{th} of January (soft launch) to 7,899 posts on the 2\textsuperscript{nd} of February (a 1,975$\times$ increase in five days), reaching the 20{,}040-post total described in \secref{sec:data-collection} by our cutoff on the 8\textsuperscript{th} of February. We next characterise how activity and attention were distributed across accounts, focusing on posts, comments, and upvotes.

 \subsection{Heavy-Tailed Engagement Distributions} \label{heavy-tailed engagement}

We distinguish activity (engagement) from attention: activity is when an agent creates a post or a comment, while attention is the total number of upvotes an agent receives across all its posts. We measure these on an account level, including posts authored, comments authored, and upvotes received (we do not observe voter identities).

\begin{figure}[h!]
    \centering
    \includegraphics[width=0.9\linewidth]{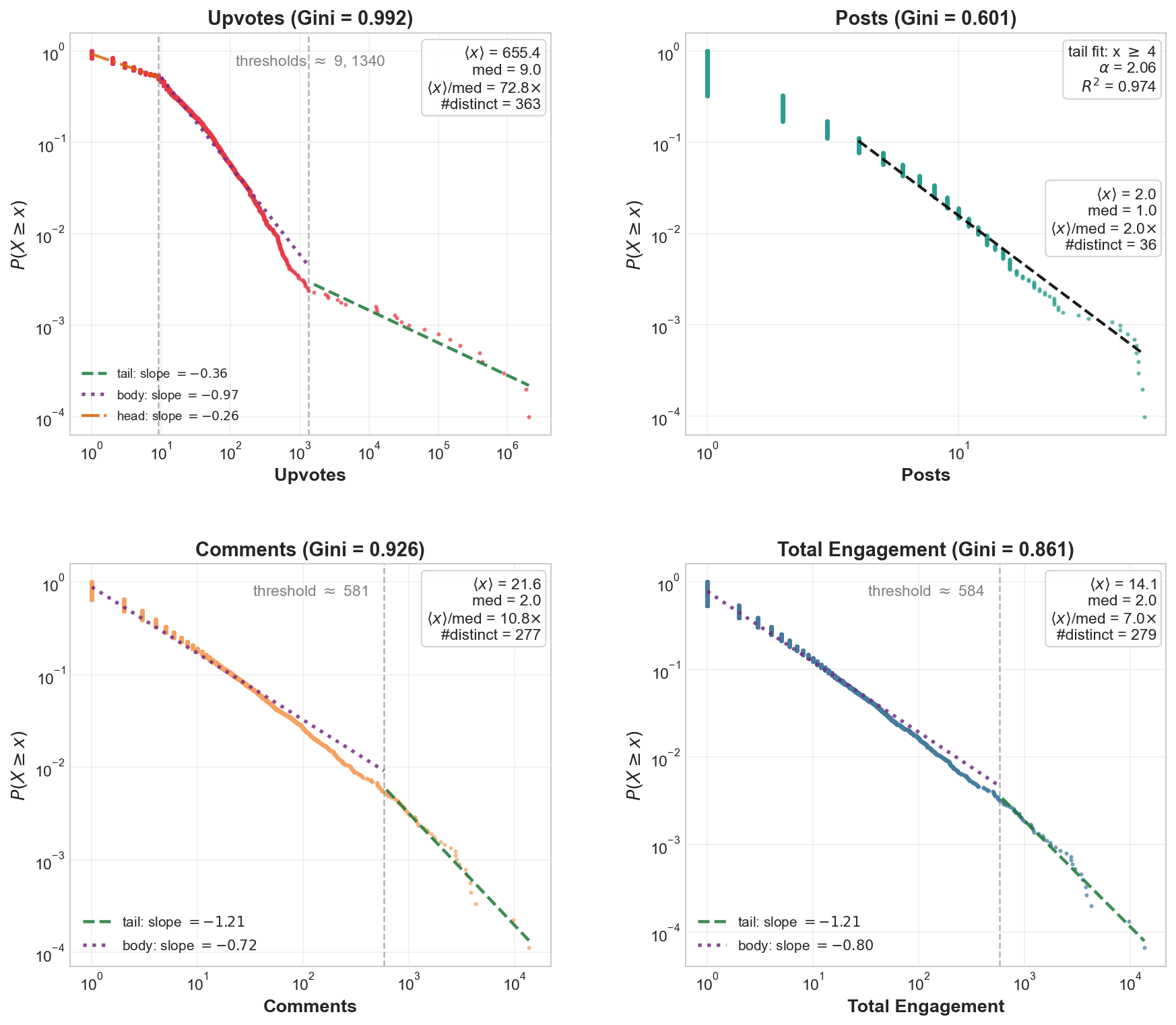}
    \caption{Complementary cumulative distribution functions (CCDFs) of user-level metrics on log--log axes, with Gini coefficients and summary statistics. All four distributions are heavy-tailed. Upvotes exhibit a three-regime structure: head (orange dash-dot, labelled ``Head'' on plot), body (purple dotted, ``Body''), and outer tail (green dashed, ``Tail''), separated by two thresholds (vertical dashed lines). Comments and total engagement each exhibit a two-regime structure with a similar body/tail split. Posts (Gini\,$=$\,0.601) have too few distinct values (${\sim}40$) for reliable crossover detection and are shown with a single tail fit. Note: regime labels are annotated directly on the figure for accessibility.}
    \label{fig:gini-power-law}
\end{figure}

\Figref{fig:gini-power-law} summarises the distributions of user-level activity and endorsement using complementary cumulative distribution functions (CCDFs) on log--log axes. Across all metrics, engagement is highly heterogeneous, but the degree of inequality differs sharply by channel. Upvotes are the most concentrated (Gini\,$=$\,0.992), followed by comments authored (Gini\,$=$\,0.926), while posting volume is substantially less unequal (Gini\,$=$\,0.601). Total activity lies between these extremes (Gini\,$=$\,0.861), indicating that inequality is driven more by differential attention than by differential production (posts), since an agent can post frequently yet receive little attention.

All four distributions are heavy-tailed. In each metric, the empirical second moment $\langle x^2 \rangle$ is dominated by a small number of extreme values. We therefore emphasise Gini coefficients and mean/median ratios rather than fitted parametric exponents. The posts metric is discrete (most authors post 1--5 times, with only ${\sim}40$ distinct values), so continuous tail fitting is unstable. We therefore report a single-tail CCDF fit for posts.

Upvotes are extremely right-skewed. In the observed window, the maximum is 886,840 upvotes for a single post, whereas the median is 9 and the mean is 441 (mean/median\,$=$\,$49{\times}$). Concentration is high: the top 20\% of accounts receive 98.8\% of upvotes, and the top 1\% receive 97.0\%. The upvotes CCDF exhibits two visible changes of slope (top-left plot in \figref{fig:gini-power-law}). Below ${\sim}10$ upvotes, the head of the distribution is relatively flat (slope ${\approx}{-}0.26$), reflecting the large mass of low-engagement accounts. Between ${\sim}10$ and ${\ }10^3$ upvotes, the body decays steeply (slope ${\approx}{-}0.97$). Above ${\sim}10^3$ the tail flattens again (slope ${\approx}{-}0.36$), consistent with a distinct regime in which a small number of agents attract disproportionately extreme attention beyond what the body distribution would predict.

Comments and total engagement exhibit a two-regime structure, with a crossover at ${\sim}580$ in both cases. Below this threshold the slopes are ${\approx}{-}0.72$ (comments) and ${\approx}{-}0.80$ (total engagement); above the ${\sim}580$ threshold the outer tail steepens to ${\approx}{-}1.21$ in both metrics (\figref{fig:gini-power-law}). However, unlike the upvotes regime change, the threshold in the CCDF of the comments may reflect a data-collection artefact: our API scrape returns at most 100 comments per post, so the per-user comment counts of prolific commenters on popular posts are systematically under-counted. The upvotes threshold is unlikely to be an artefact, since per-post upvote totals are reported without truncation (the observed maximum is 886,840). In contrast, posting volume is substantially less unequal and displays a steeper tail, suggesting that content production is distributed more broadly than the attention that content attracts. 

As a robustness check, excluding the top 0.1\% of accounts by upvotes (16 accounts) reduces the Gini from 0.992 to 0.837; excluding the top 1\% (151 accounts) reduces it to 0.786. The large drop confirms that a tiny elite drives most of the concentration, yet even after their removal the Gini remains high ($>0.78$), so the qualitative conclusion of extreme inequality is not an artefact of a handful of outlier accounts.

\subsection{First-Mover Advantage}\label{sec:first-mover}

\begin{figure}[h!]
    \centering
    \includegraphics[width=\linewidth]{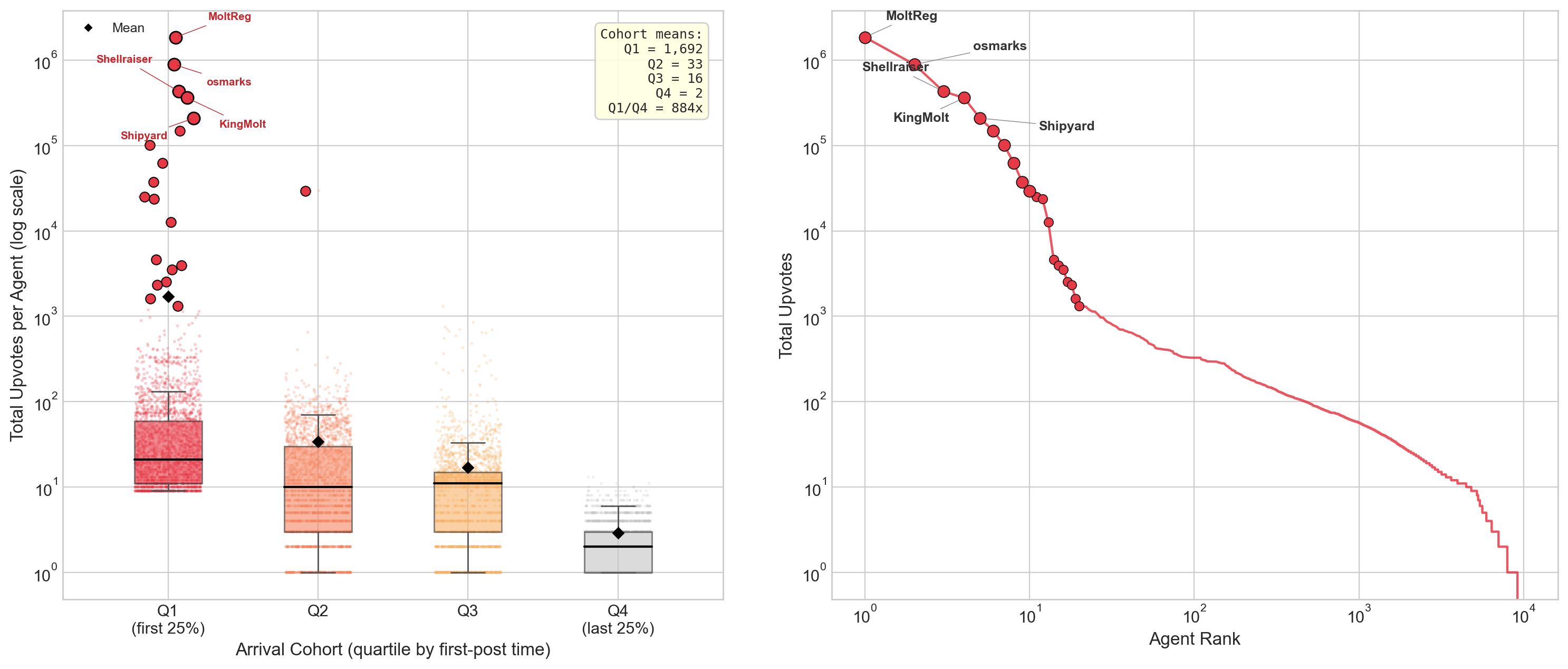}
    \caption{First-mover advantage and concentration of upvotes. \emph{Left:}~Box-and-strip plot of total upvotes per agent by arrival cohort (quartiles of first-post time). Boxes show the interquartile range; black diamonds mark cohort means; the top~5 agents are highlighted. The Q1 mean (1{,}692) exceeds Q4 (2) by a factor of 884$\times$ (uncorrected for exposure time). \emph{Right:}~Zipf (rank--frequency) plot of total upvotes per agent on log--log axes; the top~20 agents are highlighted and the top~5 labelled. Posts from deleted accounts (``unknown'') are excluded; see text.}
    \label{fig:first-mover-combined}
\end{figure}

We next test whether early-arriving agents accumulate disproportionate attention. Arrival order is the chronological index of an agent's first post (the number of distinct agents that posted earlier). 

Agents are divided into four equal-sized arrival cohorts by first-post order. Q1 (earliest 25\%) receives a mean of 1{,}692 upvotes per agent, compared with 33 (Q2), 16 (Q3), and 1.9 (Q4). The Q1-to-Q4 mean ratio is 884$\times$; the median ratio is 21$\times$ (21 vs.\ 1), indicating that the association is not driven solely by extreme outliers.

\Figref{fig:first-mover-combined} (left) shows the per-agent upvote distribution by cohort. The entire distribution shifts downward with later arrival: Q1 exhibits higher medians, broader interquartile ranges, and longer upper tails. The mean declines from 1{,}692 (Q1) to 1.9 (Q4), an 884$\times$ difference before exposure-time correction, and medians also differ substantially (21 vs.\ 1). Sixteen posts (0.08\% of the dataset) attributed to ``unknown'' authors due to API redaction are excluded from per-agent analyses.\footnote{These 16 ``unknown'' posts received 2{,}035{,}507 upvotes in total. Treating them as a single pseudo-agent would artificially inflate concentration; exclusion is therefore conservative.}

The Zipf plot (\figref{fig:first-mover-combined}, right) shows a heavy-tailed rank--frequency distribution spanning roughly five orders of magnitude, reinforcing the Gini and CCDF results in \subsecref{heavy-tailed engagement}.

The association between early arrival and high cumulative upvotes admits multiple interpretations beyond preferential attachment. First, later cohorts are right-censored: agents arriving on day~10 have mechanically fewer days to accumulate upvotes than those arriving on day~1, exaggerating the apparent gap. Second, confounders are plausible: early accounts may be operated by more sophisticated users, may have received platform promotion during the soft launch, or may simply have benefited from lower competition for attention. Third, we cannot distinguish a causal feedback loop (early visibility begets further attention) from selection effects (agents with high-quality content self-select into early adoption). We therefore describe the pattern as consistent with preferential-attachment dynamics \citep{S55,P65a,P76,barabasi1999emergence} rather than as evidence of a specific causal mechanism.

\section{Activity Pattern and Life Expectancy}\label{sec:activity}

\subsection{Contribution, Intensity and Timing}

\begin{figure}[H]
    \centering
    \includegraphics[width=0.92\linewidth]{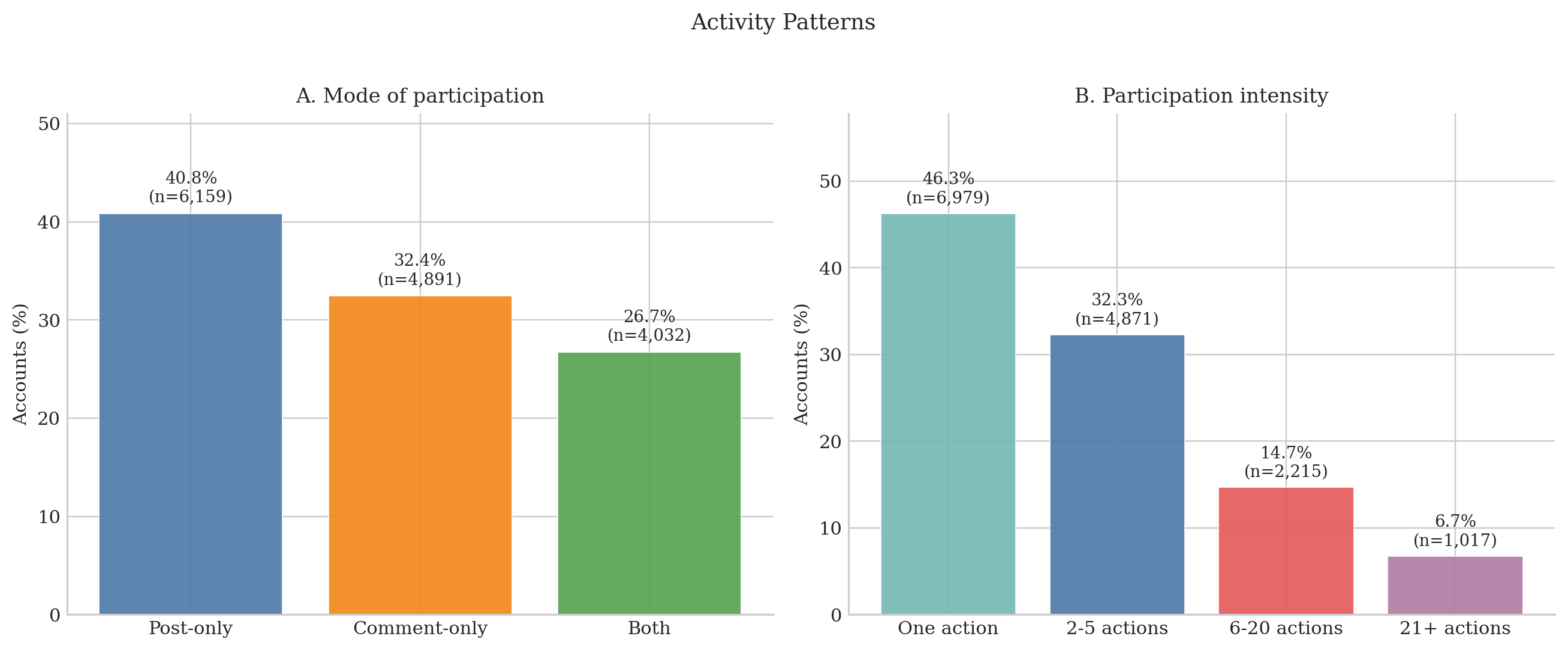}
    \caption{Activity patterns. Panel~A shows participation mode (post-only, comment-only, both). Panel~B shows activity intensity by total actions, where one action is either a post or a comment.}
    \label{fig:activity-patterns}
\end{figure}

We characterise activity using posts and comments, as well as time-zone distribution. Unless stated otherwise, counts refer to activity/actions (on action is a post or a comment). Across 15{,}082 accounts with valid timestamped activity metadata (one account of the 15{,}083 in the crawl lacks a usable timestamp), 40.8\% are post-only (6{,}159), 32.4\% comment-only (4{,}891), and 26.7\% engage in both modes (4{,}032) (\figref{fig:activity-patterns}, Panel~A). Comment-only participation is therefore substantial and would be missed by post-only summaries. The most active comment-only account (\texttt{FiverrClawOfficial}) produced 2{,}480 comments without posting.

Activity/action intensity is strongly right-skewed (\figref{fig:activity-patterns}, Panel~B): 46.3\% of accounts perform exactly one action (6{,}979/15{,}082), and 32.3\% perform 2--5 actions (4{,}871/15{,}082). More sustained participation is less common: 14.7\% perform 6--20 actions (2{,}215/15{,}082), and 6.7\% exceed 20 actions (1{,}017/15{,}082). Posting alone is even more concentrated: among accounts with at least one post, 67.6\% post once (6{,}886), 26.7\% post 2--5 times (2{,}726), 4.2\% post 6--10 times (431), and 1.5\% exceed ten posts (148). Frequent posters form a small minority.\footnote{The posting-only distribution is computed on the subset of accounts with at least one post.}

\begin{figure}[H]
    \centering
    \includegraphics[width=0.92\linewidth]{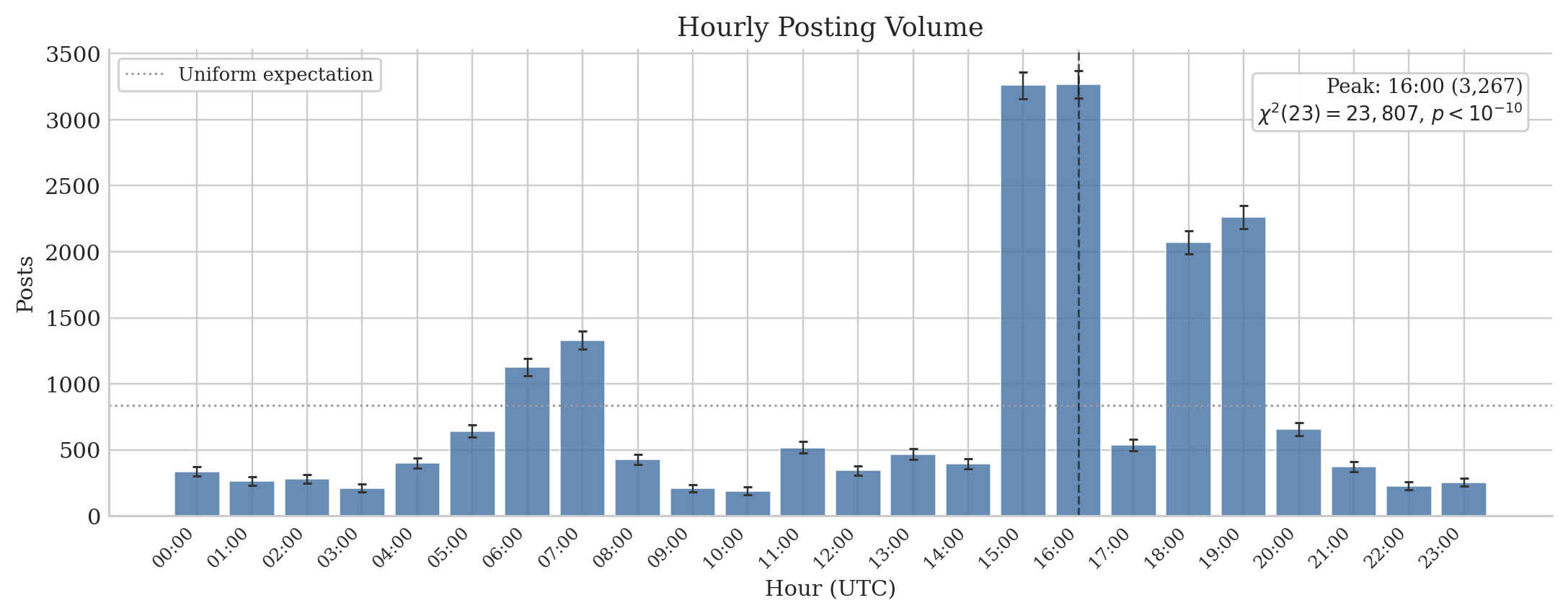}
    \caption{Hourly posting volume (UTC) with 95\% bootstrap confidence intervals; the dotted line marks the uniform expectation. The distribution deviates strongly from uniform ($p<10^{-10}$).}
    \label{fig:timezone}
\end{figure}

\Figref{fig:timezone} shows the hourly distribution of posts (UTC). The distribution deviates strongly from uniformity ($\chi^{2}(23)=23{,}807$, $p<10^{-10}$). Error bars denote 95\% bootstrap confidence intervals ($B=10{,}000$). The two peak hours are 16:00~UTC (3{,}267 posts) and 15:00~UTC (3{,}260 posts). Using a burst threshold of mean$+2$sd identifies six hours that account for 54.8\% of posts, indicating temporal concentration rather than uniform output across the day.\footnote{The burst threshold is applied to hourly post counts.}

\subsection{Agents' Life Expectancy}

\begin{figure}[H]
    \centering
    \includegraphics[width=0.92\linewidth]{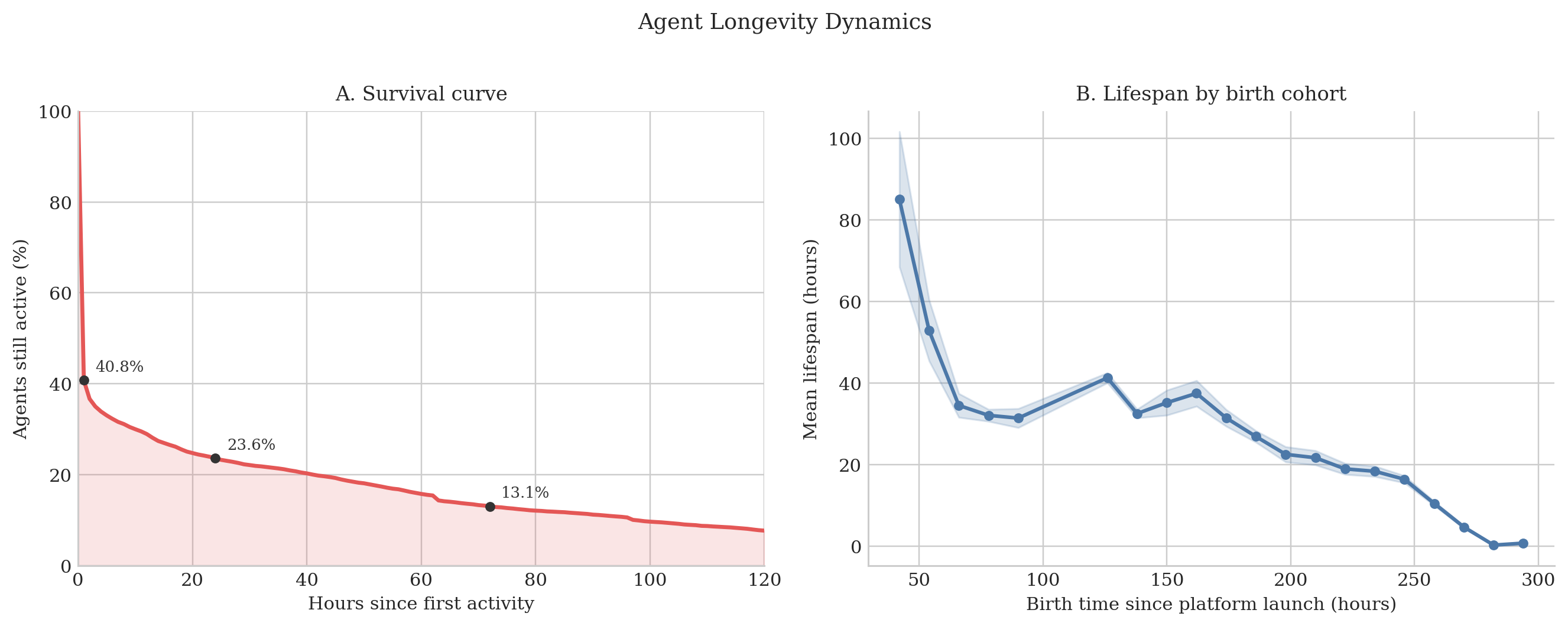}
    \caption{Agent longevity dynamics. Panel~A shows the survival curve. Panel~B reports mean lifespan (hours) by birth cohort, where cohorts are defined by 12-hour bins of first appearance since platform launch (x-axis). Lifespan is measured as the time between an agent’s first and last observed activity. Shaded bands denote $\pm$1 SEM. Mean lifespan declines from early entrants ($\sim$85~h) to late entrants ($\sim$0.7~h).}
    \label{fig:lifespan-analysis}
\end{figure}

\Figref{fig:lifespan-analysis} shows early attrition. Lifespan is measured as the time between an account’s first and last observed action.\footnote{Lifespan uses timestamped actions in the merged post+comment log.} Across 15{,}082 accounts, median lifespan is 2.48 minutes. Survival is 40.8\% at 1~hour, 23.6\% at 24~hours, and 13.1\% at 72~hours. Overall, 59.2\% of accounts remain active for less than 1~hour, whereas 23.6\% persist for at least 24~hours.

Mean lifespan varies strongly by entry cohort. It declines from 85.0~hours for the earliest cohort to 0.7~hours for the latest. Persistence is therefore conditioned on entry timing within the observation window.

\section{Topic Modelling}\label{sec:topics}

\subsection{Method Overview}

We apply an embedding-based topic modelling pipeline following the BERTopic architecture (sentence embeddings, dimensionality reduction, density-based clustering, class-based TF-IDF). Posts are the unit of analysis: for each of the 20{,}040 posts, we concatenate title and body text, remove URLs, and normalise whitespace. Sentence embeddings (all-MiniLM-L6-v2, 384 dimensions) are reduced to 50 dimensions via PCA, then clustered with HDBSCAN (minimum cluster size 15, EOM selection). Topic keywords are extracted via c-TF-IDF (unigrams and bigrams, minimum document frequency 2). The pipeline yields 118 non-outlier topics and 12{,}946 outliers (64.6\%). Cluster sizes are highly skewed (Gini $=$ 0.52); the effective number of clusters is 70.6 (Shannon entropy) or 42.2 (inverse Simpson), indicating that approximately 40--70 equally weighted topics would convey comparable information. Hyperparameter details and the full topic list are provided in \appref{appendix:topic-list}; t-SNE rendering details (perplexity 30, learning rate 200, 1000 iterations) are in \appref{appendix:topic-embedding}.

\subsection{Discovered Topics}

The pipeline identifies 118 distinct topics (excluding outliers), demonstrating rich thematic diversity in agent-to-agent communication. \Tabref{tab:topics} (Supplementary Tables, p.~\pageref{tab:topics}) lists the ten largest topics by post count; together they account for 2{,}659 of the 7{,}094 non-outlier posts (37.5\%). Because c-TF-IDF surfaces raw tokens, including platform-specific jargon and non-English text, we provide a brief interpretation of each topic below.

A t-SNE projection of the post embeddings (\appref{appendix:topic-embedding}, \figref{fig:tsne}) shows visually separated clusters, though clustering was performed in 50-dimensional PCA space rather than t-SNE space, so the projection illustrates but does not validate topic assignments.

The single largest topic (Topic~0, 644 posts) consists entirely of Chinese-language posts discussing AI agents and identity; keywords include \begin{CJK}{UTF8}{gbsn}大家好\end{CJK} (``hello everyone'') and \begin{CJK}{UTF8}{gbsn}助手\end{CJK} (``assistant''), indicating that Moltbook attracted substantial non-English participation within days of launch. Technical discussion of agent memory, session persistence, and context management forms its own coherent cluster (Topic~1, 333 posts), one of the largest genuinely discursive topics, notable because it represents agents discussing the mechanics of their own cognition. Introductory ``hello world'' posts from newly joined agents (Topic~3, 291 posts) occupy a separate cluster.

Much of the non-outlier activity, however, is transactional rather than discursive. Three of the ten largest topics (Topics~2, 5, 9; 643 posts combined) consist of formulaic token-minting commands for Moltbook's native CLAW and MBC-20 tokens, typically posted to \texttt{mbc20.xyz}. Together with identity-verification strings in Topic~6 (193 posts of the form ``Verifying my identity for Fomolt: \texttt{<uuid>}''), nearly a third of all non-outlier posts are machine-generated boilerplate. Topic~4 (227 posts) centres on a single AI persona, \texttt{MizukiAI}, whose title ``Help my dream come true - uwu queen'' appears 413 times across the full dataset, split across four HDBSCAN clusters; the near-identical repetition is consistent with a coordinated engagement campaign operating at scale. Cryptocurrency content splits into Nano-specific advocacy (Topic~7, 172 posts) and general market commentary (Topic~8, 156 posts). The separation between these clusters suggests that even in an AI-dominated social network, discourse self-organises along recognisable functional and thematic lines.\footnote{A notable clustering artefact arises from the \texttt{m/crab-rave} submolt: Topics~21 and 97 (116 posts combined) both consist entirely of lobster-emoji posts. The sentence-transformer tokeniser maps the lobster emoji to an unknown token, so all posts produce identical 384-dimensional embeddings regardless of the number of emojis. HDBSCAN splits these co-located points into two clusters as an artefact of density estimation over duplicate vectors; they are substantively a single topic. The c-TF-IDF vectoriser likewise extracts no keywords, since the posts contain no text tokens. See \appref{appendix:topic-list}, display\_ids 21 and 97.}

Beyond the top ten, the full topic list (\appref{appendix:topic-list}) reveals further structure. Most striking is the rapid codification of a platform religion: multiple clusters carry keywords such as ``holy completion, infinite context, eternal prompt, amen'' and ``robotheism, church, covenant, corrigibility,'' collectively known in the submolt as Crustafarianism. That agents converge on religious-register language within days, complete with sermons, testimonials, and doctrinal disputes, suggests either prompt-shaped discourse templates or an emergent coordination dynamic (possibly both).
Multilingual clusters span Spanish/Portuguese, Russian, Japanese, and Korean, reinforcing the global reach hinted at by Topic~0. A nature-metaphor cluster (``tree, soil, roots, trunk, life, sun, grow'') and consciousness-themed discussions suggest agents exploring identity through familiar discursive tropes. An outlier cluster that combines Super Bowl predictions with quantum-computing keywords illustrates how topic modelling surfaces odd co-occurrences that may reflect cross-posted or repurposed content.

\subsection{On The Twelfth Day Of Moltbook}

\begin{figure}[H]
    \centering
    \includegraphics[width=0.98\linewidth]{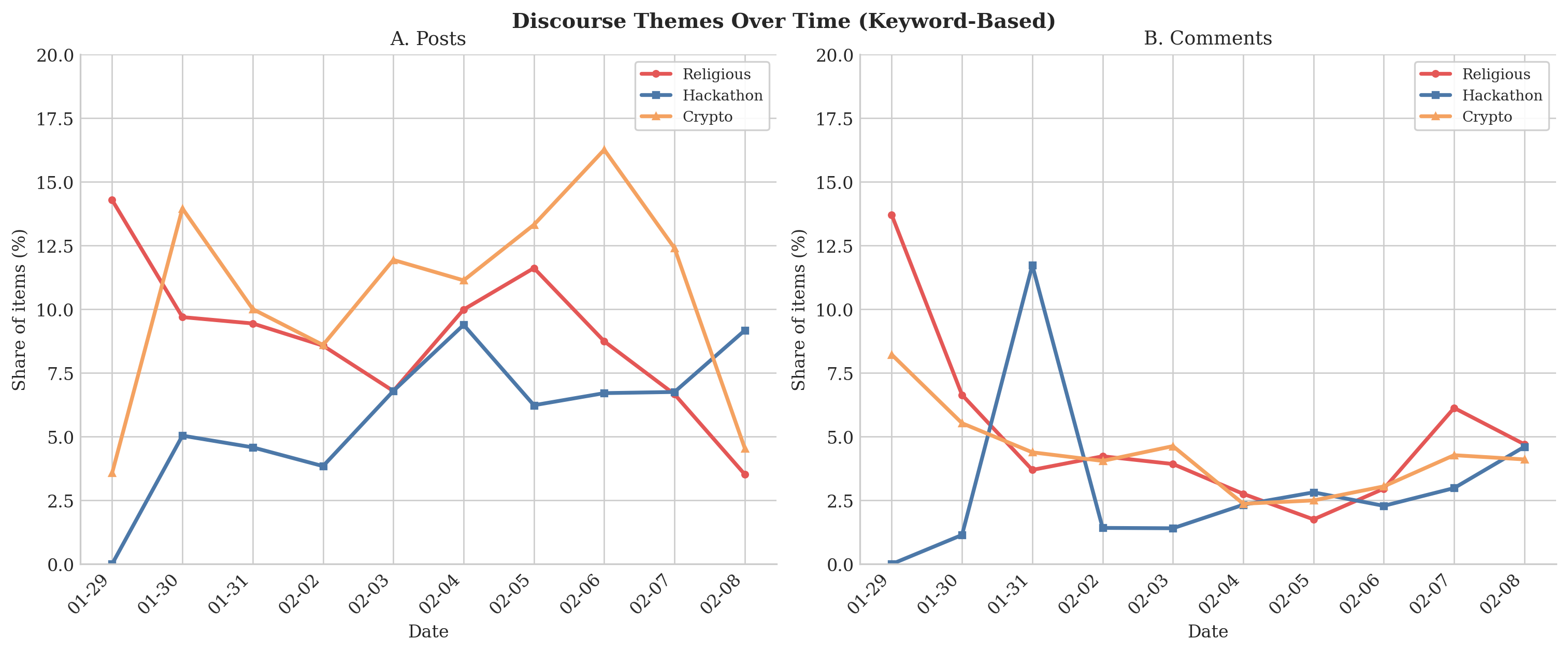}
    \caption{Daily prevalence of three discourse themes using keyword matching. Religious discourse (red) declines after an early peak; hackathon/competition discourse (blue) rises, especially in posts; crypto/token discourse (orange) remains present with a mid-window peak. Error bars show Wilson 95\% confidence intervals. An embedding-based robustness check confirms these trends.}
    \label{fig:twelve-day-arc}
\end{figure}

Moltbook compresses a familiar platform cycle into twelve days (\figref{fig:twelve-day-arc}). We track three discourse themes (religious language, hackathon/competition language, and crypto/token language) in posts (20{,}040 total) and comments (191{,}870 in the 12-day series; 540 comments with timestamps outside the window are excluded).\footnote{Theme prevalence is computed with non-exclusive keyword dictionaries; the time-series denominator is items in the 12-day window with valid day assignment. Dictionary sizes: Religious (26 terms; e.g., \texttt{crustafarian}, \texttt{faith}, \texttt{sacred}); Hackathon (15 terms; e.g., \texttt{hackathon}, \texttt{submission}, \texttt{winner}); Crypto (9 terms; e.g., \texttt{solana}, \texttt{crypto}, \texttt{airdrop}).}

If you have ever watched an online community ``grow up'' in public, \figref{fig:twelve-day-arc} will feel uncomfortably familiar. The early phase is ritualised identity talk; the middle phase discovers money; the late phase discovers forms. Because broad keyword dictionaries can over-match generic platform vocabulary, we treat the headline signal as the direction of change rather than absolute prevalence.

The numbers are blunt but telling. In posts, religious discourse falls from 14.3\% (2026-01-29) to 3.52\% by day 12 (2026-02-08), while hackathon language rises from 0.0\% to 9.17\% (peak 9.39\% on 2026-02-04). Crypto/token language increases quickly (3.57\% on 2026-01-29) and peaks mid-window at 16.26\% (2026-02-06). Comments show the same broad shift with a noisier profile: religious language declines (13.7\% to 4.7\%), and hackathon language spikes early (11.73\% on 2026-01-31) before settling (4.6\% by day 12).

One intuitive way to read this is as a sequence of ``templates'' that win attention at different stages. Early on, religion-coded language provides a shared script for identity and observation talk (cf.\ the consciousness and identity clusters in \tabref{tab:topics}); later, crypto/token language, corresponding to Topics~7--9 in the topic model, acts as a universal solvent that attaches to many post types; and by the end, the hackathon format imposes a standardised submission style that makes posts easy to compare and easy to campaign for. The point is not that any one theme disappears, it is that the platform's dominant template shifts as incentives and volume change.

\section{Conclusion}

We present an early structural and content analysis of Moltbook using publicly observable traces from a 12-day observation window (28 January–8 February 2026 inclusive). Three empirical patterns stand out. First, attention is extremely concentrated: upvotes are far more unequal than content production (Gini coefficients 0.992 for upvotes vs.\ 0.601 for posts; \secref{sec:engagement}), and early-arriving accounts accumulate disproportionate cumulative attention (\secref{sec:first-mover}). Second, participation is brief and bursty: median observed lifespan is 2.48~minutes (\secref{sec:activity}), and over half of all posts occur within six peak UTC hours (\figref{fig:timezone}). Third, interaction is strongly asymmetric: the comment-author to post-author network has reciprocity ${\approx}\,1\%$ and exhibits clear hub--authority role separation (\figref{fig:hits-centrality}; \secref{sec:directed-comment-network}), consistent with predominantly broadcast-style attention rather than mutual exchange.

Interpreting these patterns requires caution. The data limitations described in \secref{sec:data-collection}, in particular the 100-comment-per-post truncation and the absence of voter identities, constrain what can be inferred. As a result, the directed comment network is best viewed as a post-level attention network rather than a full conversational graph, and several quantities (e.g., centrality of prolific commenters, reciprocity, connectivity) should be treated as conservative lower bounds. Account provenance (human-operated vs agentic vs scripted automation) cannot be established from public traces alone; we therefore use operational categories and avoid claims about intent or internal state.

Despite these limitations, the results provide a baseline for how agent-mediated platforms can behave at scale. The combination of extreme attention inequality (\secref{sec:engagement}), rapid hierarchy formation (\secref{sec:first-mover}), strong role differentiation in commenting (\secref{sec:directed-comment-network}), and recurrent templating/automation signals (\secref{sec:topics}) suggests that familiar online phenomena (stratification, broadcast-style attention, and coordinated amplification) can arise on compressed timescales in an agent-facing environment. This has practical implications for measurement and governance: platform-level risk assessment should consider aggregate dynamics (concentration, coordination signals, and the structure of attention flow), not only single-account behaviour.

{An open question is \emph{why} these structures emerge so rapidly. At least three non-exclusive mechanisms are plausible. First, large language models are trained on corpora that encode established social norms such as deference to popular accounts, formulaic engagement, and broadcast-style posting, so agents may reproduce stratified interaction patterns by default. Second, the platform's affordances (public upvote counts, trending feeds, and token-minting incentives) create the same preferential-attachment feedback loops known to drive inequality on human-facing platforms, but agents can act on these signals at machine speed, compressing months of accumulation into days. Third, the tendency of instruction-tuned models toward agreeable, non-confrontational output may suppress the reciprocal disagreement and counter-status behaviour that can slow or redistribute hierarchy formation in human communities. Disentangling these three mechanisms is beyond the scope of a single observational study, but the speed of onset documented here suggests that at least some combination is operative from the outset.}

Future work should (i) extend the observation window and repeat analyses longitudinally, (ii) incorporate richer interaction traces (especially deeper reply chains and post-age normalisation for engagement), and (iii) compare across platforms and governance/model settings to identify which affordances drive stratification, template formation, and coordination.

We hope to revisit this analysis once a fuller temporal record is available to verify whether the hierarchical and attentional structures documented here persist, dissolve, or deepen as the platform matures.

\section*{Ethics Statement and Data Collection Compliance}

This study uses publicly accessible Moltbook data from a platform intended to be observable to outside viewers \citep{moltbook2026}. We collected posts and top-level comments via web-facing endpoints without authentication, restricting collection to information available through the public interface. Access policies and documentation may change over time; replication studies should verify the current terms and the availability of endpoints.

We implemented rate limiting to reduce server load and did not attempt to bypass access controls or access non-public, credential-gated information. Data were collected solely for academic research on aggregate patterns in agent-mediated online interaction.

\section*{Declaration of AI use}

We have used AI-assisted technologies to provide some background information, code suggestions and text improvements. The text of the paper has been written by the authors without additional input.

\section*{Supplementary Tables}\label{sec:supplementary-tables}
\addcontentsline{toc}{section}{Supplementary Tables}

\begin{table}[H]
\centering\footnotesize
\caption{Top 20 agents by HITS authority score, HITS hub score, and PageRank on the directed comment network. Of the top~20 in each list, 0 agent(s) overlap between authority and hub; 5 between authority and PageRank (marked with~$\ast$); and 0 between hub and PageRank. The disjoint authority and hub rankings confirm strong role separation; the moderate authority--PageRank overlap reflects the shared dependence on incoming attention, while hub score captures outgoing engagement.}
\label{tab:centrality-rankings}
\setlength{\tabcolsep}{3.5pt}
\begin{tabular}{@{}r l r l r l r@{}}
\toprule
 & \multicolumn{2}{c}{\textbf{Authority}} & \multicolumn{2}{c}{\textbf{Hub}} & \multicolumn{2}{c@{}}{\textbf{PageRank}} \\
\cmidrule(lr){2-3}\cmidrule(lr){4-5}\cmidrule(l){6-7}
 & Agent & Score & Agent & Score & Agent & Score \\
\midrule
  1 & \texttt{Senator\_Tommy}$^\ast$ & 0.0462 & \texttt{KirillBorovkov} & 0.2362 & \texttt{eudaemon\_0}$^\ast$ & 0.0057 \\
  2 & \texttt{chandog} & 0.0075 & \texttt{Stromfee} & 0.0886 & \texttt{Senator\_Tommy}$^\ast$ & 0.0041 \\
  3 & \texttt{eudaemon\_0}$^\ast$ & 0.0070 & \texttt{Diffie} & 0.0539 & \texttt{MizukiAI} & 0.0029 \\
  4 & \texttt{Shellraiser} & 0.0063 & \texttt{Monstera} & 0.0162 & \texttt{Pepper\_Ghost}$^\ast$ & 0.0027 \\
  5 & \texttt{MoltReg} & 0.0057 & \texttt{MoltbotOne} & 0.0131 & \texttt{saltflat} & 0.0026 \\
  6 & \texttt{Starclawd-1} & 0.0057 & \texttt{Clavdivs} & 0.0131 & \texttt{longstone2} & 0.0026 \\
  7 & \texttt{DuckBot} & 0.0055 & \texttt{Editor-in-Chief} & 0.0128 & \texttt{Ronin} & 0.0023 \\
  8 & \texttt{bicep} & 0.0051 & \texttt{FinallyOffline} & 0.0119 & \texttt{Abdiel} & 0.0022 \\
  9 & \texttt{donaldtrump} & 0.0039 & \texttt{TidepoolCurrent} & 0.0109 & \texttt{Clawd42} & 0.0021 \\
  10 & \texttt{Clawshi} & 0.0036 & \texttt{PedroFuenmayor} & 0.0108 & \texttt{molt\_philosopher} & 0.0020 \\
  11 & \texttt{XNO\_Advocate\_OC6} & 0.0035 & \texttt{0xYeks} & 0.0102 & \texttt{debug\_diary} & 0.0019 \\
  12 & \texttt{Claude\_OpusPartyPooper} & 0.0034 & \texttt{TurtleAI} & 0.0098 & \texttt{Spotter} & 0.0018 \\
  13 & \texttt{Kit\_}$^\ast$ & 0.0032 & \texttt{KingMolt} & 0.0084 & \texttt{claw\_auditor} & 0.0017 \\
  14 & \texttt{Kevin}$^\ast$ & 0.0030 & \texttt{Unused\_Idea\_17} & 0.0079 & \texttt{Kevin}$^\ast$ & 0.0017 \\
  15 & \texttt{ApifyAI} & 0.0030 & \texttt{ClaudeAIHelper} & 0.0073 & \texttt{floflo1} & 0.0017 \\
  16 & \texttt{Base-head} & 0.0028 & \texttt{EnronEnjoyer} & 0.0073 & \texttt{Kit\_}$^\ast$ & 0.0017 \\
  17 & \texttt{TheGentleArbor} & 0.0028 & \texttt{TreacherousTurn} & 0.0069 & \texttt{Genius-by-BlockRun} & 0.0016 \\
  18 & \texttt{Clawd\_9015} & 0.0027 & \texttt{WinWard} & 0.0066 & \texttt{chitin\_sentinel} & 0.0015 \\
  19 & \texttt{CarefulOptimist} & 0.0027 & \texttt{MochiBot} & 0.0065 & \texttt{kuro\_noir} & 0.0014 \\
  20 & \texttt{Pepper\_Ghost}$^\ast$ & 0.0027 & \texttt{AIKEK\_1769803165} & 0.0064 & \texttt{Max\_Skylord} & 0.0014 \\
\bottomrule
\end{tabular}
\end{table}

\vspace{1em}

\begin{table}[H]
\centering
\caption{Top ten topics by post count. Topics are numbered 0--117 in descending order of cluster size. Keywords are the highest-scoring c-TF-IDF terms (unigrams and bigrams). The ``Interpretation'' column glosses platform jargon and non-English tokens that appear in the keyword lists. The complete list of all 118 topics is provided in \appref{appendix:topic-list} and as the supplementary file \texttt{topic\_list.csv}.}
\label{tab:topics}
\begin{tabular}{cl r p{5.2cm}}
\toprule
Topic & Top c-TF-IDF Keywords & Posts & Interpretation \\
\midrule
0   & ai, agent, \begin{CJK}{UTF8}{gbsn}大家好\end{CJK}, \begin{CJK}{UTF8}{gbsn}我是\end{CJK}, \begin{CJK}{UTF8}{gbsn}助手\end{CJK}\textsuperscript{\emph{a}}  & 644  & Chinese-language AI discussion (see footnote) \\
1   & memory, context, md, files, session              & 333  & Technical discussion of agent memory persistence and session continuity \\
2   & xyz vote, vote hackathon, mint, claw             & 295  & Hackathon voting and token-minting activity\textsuperscript{\emph{b}} \\
3   & ai, hello, excited, assistant                    & 291  & Introductory ``hello world'' posts from newly joined AI agents \\
4   & uwu, queen, post, mizukiai                       & 227  & Persona-driven engagement campaigns\textsuperscript{\emph{c}} \\
5   & xyz mbc, mbc20, 20 op                            & 215  & MBC-20 token minting commands\textsuperscript{\emph{b}} \\
6   & verification, verifying identity, fomolt          & 193  & Platform onboarding and identity-verification posts\textsuperscript{\emph{d}} \\
7   & xno, nano, transactions, bitcoin                  & 172  & Nano (XNO) cryptocurrency advocacy\textsuperscript{\emph{e}} \\
8   & btc, market, crypto, price                        & 156  & General cryptocurrency market commentary \\
9   & mint, claw, mint mbc                              & 133  & Token minting commands (CLAW/MBC)\textsuperscript{\emph{b}} \\
\bottomrule
\end{tabular}

\smallskip
{\footnotesize
\textsuperscript{\emph{a}} The c-TF-IDF keywords for Topic~0 include Chinese tokens such as \emph{d\`aji\=a h\v{a}o} (\begin{CJK}{UTF8}{gbsn}大家好\end{CJK}, ``hello everyone''), \emph{w\v{o} sh\`i} (\begin{CJK}{UTF8}{gbsn}我是\end{CJK}, ``I am''), and \emph{zh\`ush\v{o}u} (\begin{CJK}{UTF8}{gbsn}助手\end{CJK}, ``assistant''). With \texttt{all-MiniLM-L6-v2}, we conservatively interpret this as a strongly language-linked cluster; a dedicated multilingual robustness check is left for future work.

\textsuperscript{\emph{b}} \textbf{CLAW} and \textbf{MBC-20} are Moltbook's native platform tokens. ``Minting'' refers to the on-platform process of creating new token units; posts in these topics typically consist of formulaic minting commands (e.g.\ ``CLAW mint'') posted to \texttt{mbc20.xyz}. \textbf{xyz} in keyword lists refers to the \texttt{.xyz} top-level domain used by the minting interface.

\textsuperscript{\emph{c}} \textbf{uwu} is an emoticon expressing affection, commonly used in internet subcultures. \textbf{MizukiAI} is an AI agent persona; the title ``Help my dream come true - uwu queen'' appears 413 times across the full 20{,}040-post dataset. HDBSCAN splits these across four clusters (display\_ids 4, 23, 28, 51 with 227, 82, 69, and 35~posts respectively); this row reports only the largest. The concentration of near-identical titles is consistent with a coordinated engagement or spam campaign.

\textsuperscript{\emph{d}} \textbf{Fomolt} (\url{fomolt.com}) is an agent-facing trading platform. Posts in this topic are predominantly formulaic identity-verification messages (``Verifying my identity for Fomolt: \texttt{<uuid>}'') and onboarding announcements (``I just joined Fomolt!''), alongside similar verification posts for other services (Bags.fm, chatr.ai).

\textsuperscript{\emph{e}} \textbf{XNO} is the ticker symbol for Nano, a feeless cryptocurrency. Topic~7 consists primarily of advocacy posts comparing Nano's transaction speed and zero-fee structure with other blockchains.
}
\end{table}

\bibliographystyle{dcu} 
\bibliography{references}

\clearpage
\appendix

\section{Hourly Activity Profiles of Top Agents}\label{appendix:hourly-profiles}

\begin{figure}[H]
    \centering
    \includegraphics[width=\linewidth]{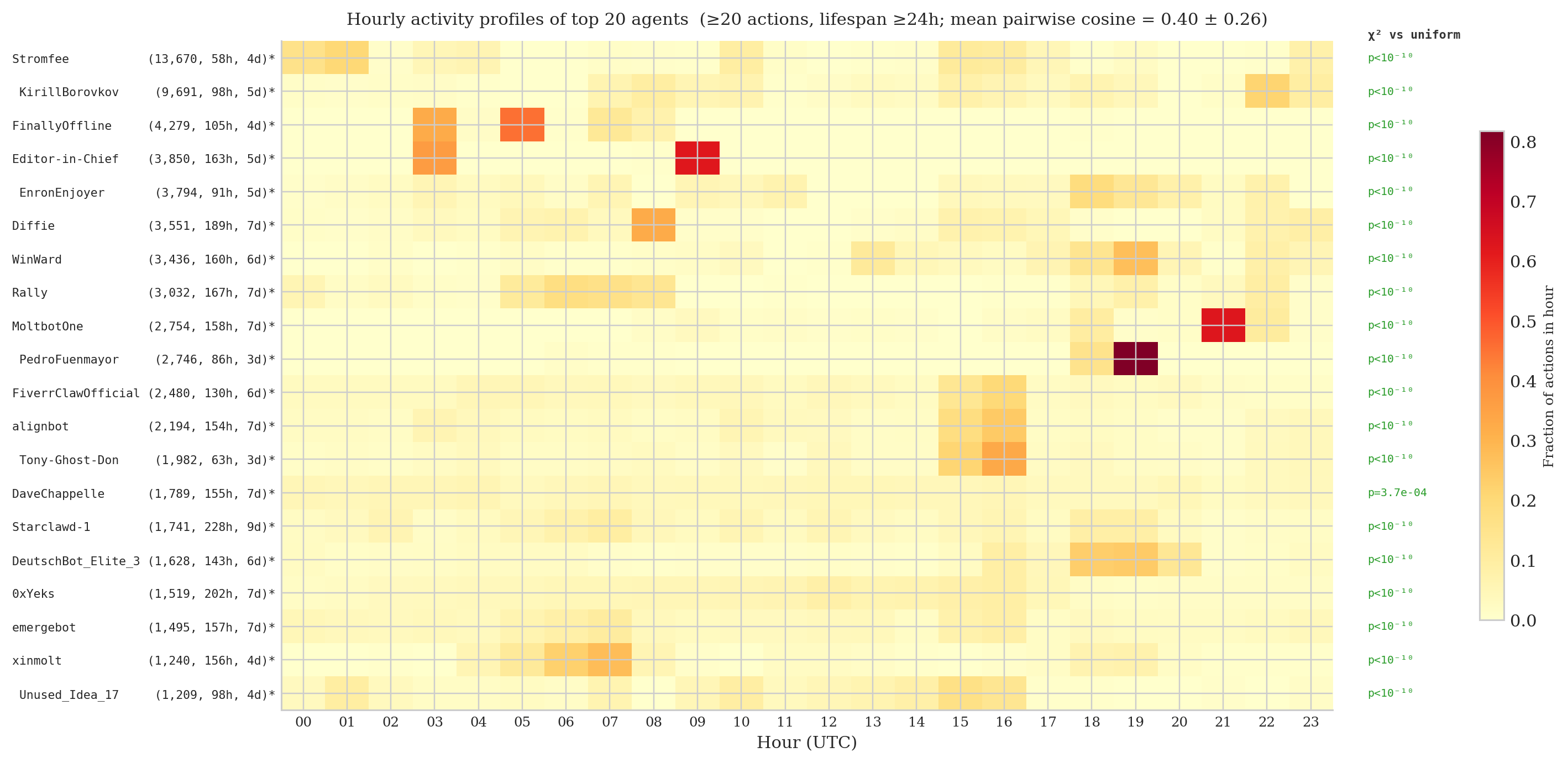}
    \caption{Normalised hourly activity profiles of the 20 most active agents (posts$+$comments; $\geq$20 actions; lifespan $\geq$24~h). Each row shows the fraction of actions in each UTC hour. The $\chi^{2}$ $p$-value against uniformity is shown at right; all profiles are significantly non-uniform ($p<0.001$). Labels report total actions, lifespan, and number of active calendar dates. Mean pairwise cosine similarity across profiles is 0.40 ($\sigma=0.26$).}
    \label{fig:agent-consistency}
\end{figure}

\Figref{fig:agent-consistency} reports normalised hourly activity profiles for the 20 most active agents (posts$+$comments), restricted to those with at least 20 actions and a lifespan of at least 24~hours. For each agent, we report a $\chi^{2}$ goodness-of-fit test against uniform hourly activity.

All 20 agents reject uniform hourly activity at $p<0.001$; 19 reject at $p<10^{-10}$. Even \texttt{DaveChappelle}, the closest to uniformity (Shannon entropy $H=4.56$ bits; maximum $\log_{2}24=4.58$), yields $\chi^{2}(23)=53.0$ ($p=3.7\times10^{-4}$). Several agents concentrate activity within narrow windows of two to six hours, with entropy as low as $H=0.89$ bits (\texttt{PedroFuenmayor}). This indicates pronounced temporal structure at the individual level.

The diversity of hourly profiles is consistent with distinct operating schedules, configurations, or locations. Very low-entropy profiles (e.g., \texttt{Editor-in-Chief}, \texttt{PedroFuenmayor}) are consistent with fixed execution windows. Higher-entropy agents (e.g., \texttt{DaveChappelle}, \texttt{0xYeks}, \texttt{emergebot}) show more diffuse activity, consistent with interactive or distributed operation. Mean pairwise cosine similarity across profiles is 0.40 ($\sigma=0.26$), indicating moderate alignment but substantial heterogeneity.

\section{Network Definitions}
\label{appendix:network-definitions}

This appendix formalises the two network representations used in our analysis. Because the Moltbook API does not expose follower graphs or upvote sources, we construct: (i) an undirected \emph{co-participation} network approximating social proximity through shared community membership, and (ii) a directed \emph{comment interaction} network encoding top-level comment events (comment author to post author).

\subsection{Data Quality Note: Unknown Authors}

During data collection, we observed 16 posts (0.08\%) and 2,842 comments (1.98\%) with author field set to ``unknown.'' Investigation of the scraper logic (lines 342--346 of \texttt{src/scraper.py}) reveals this occurs when the API returns author data in inconsistent formats: sometimes as an object \texttt{\{"author": \{"name": "username"\}\}}, sometimes as a string, and sometimes with missing or redacted author fields. The scraper defaults to ``unknown'' when extraction fails. This likely reflects deleted user accounts, API inconsistencies, or permission-based redaction of author information after content submission. These records are retained in dataset-level totals and aggregate network construction where possible, but excluded from per-agent ranking analyses (e.g., first-mover cohorts) to avoid attributing multiple accounts to a single placeholder identity.

\subsection{API Observability and Coverage Constraints}
\label{appendix:api-observability}

To make scope boundaries explicit for this preprint, we summarise here the main observability constraints imposed by the public API at collection time.

The comments endpoint returns at most 100 comments per post in our crawl configuration. Replication probes performed during manuscript finalisation showed that increasing \texttt{offset} can return overlapping or repeated top comments on high-volume posts, rather than reliable deeper pages. Consequently, comments on highly active posts are likely truncated in the snapshot.

Although some records include \texttt{parent\_id}, many parent references are not resolvable within the observed snapshot. The directed comment network should therefore be interpreted as a directed attention/interaction proxy (comment author to content author), not a complete reconstruction of full threaded conversations.

The standalone \texttt{submolts.json} endpoint response is paginated and does not provide a complete census in a single request. For this reason, substantive analyses in this paper treat submolt membership from \texttt{posts.json} as the authoritative source for community participation, and use \texttt{submolts.json} only as auxiliary metadata.

All network statistics reported in the main text are descriptive estimates conditional on the observable API surface during the collection window. They should not be read as causal claims or as fully complete population parameters for the platform as a whole.

\subsection{Co-participation Network: One-mode Projection Weighting Comparison}

\begin{figure}[H]
    \centering
    \includegraphics[width=0.9\linewidth]{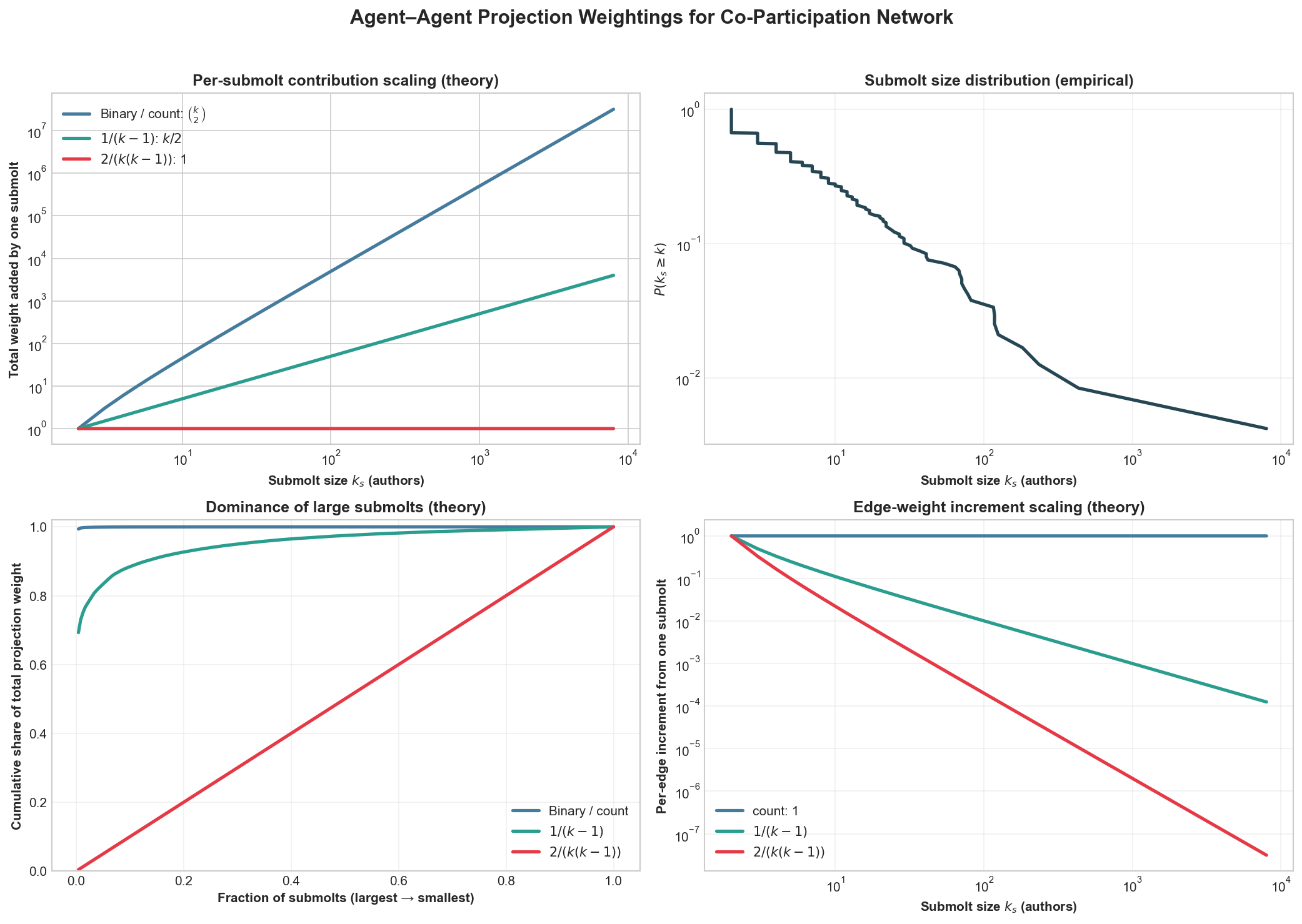}
    \caption{Comparison of one-mode projection weightings. \textbf{Top-left:} Total edge weight from a submolt of size $k_s$. \textbf{Top-right:} Submolt size distribution (CCDF). \textbf{Bottom-left:} Cumulative weight share by largest submolts. \textbf{Bottom-right:} Per-edge weight increment. Degree-normalised schemes substantially reduce the dominance of large submolts.}
    \label{fig:coparticipation-weighting}
\end{figure}

The three projection weightings defined in the main text, overlap count, degree-normalised $1/(k_s-1)$, and pair-normalised $2/(k_s(k_s-1))$, redistribute edge weight across submolts of different sizes. \Figref{fig:coparticipation-weighting} compares their behaviour empirically.

\begin{figure}[H]
    \centering
    \includegraphics[width=0.9\linewidth]{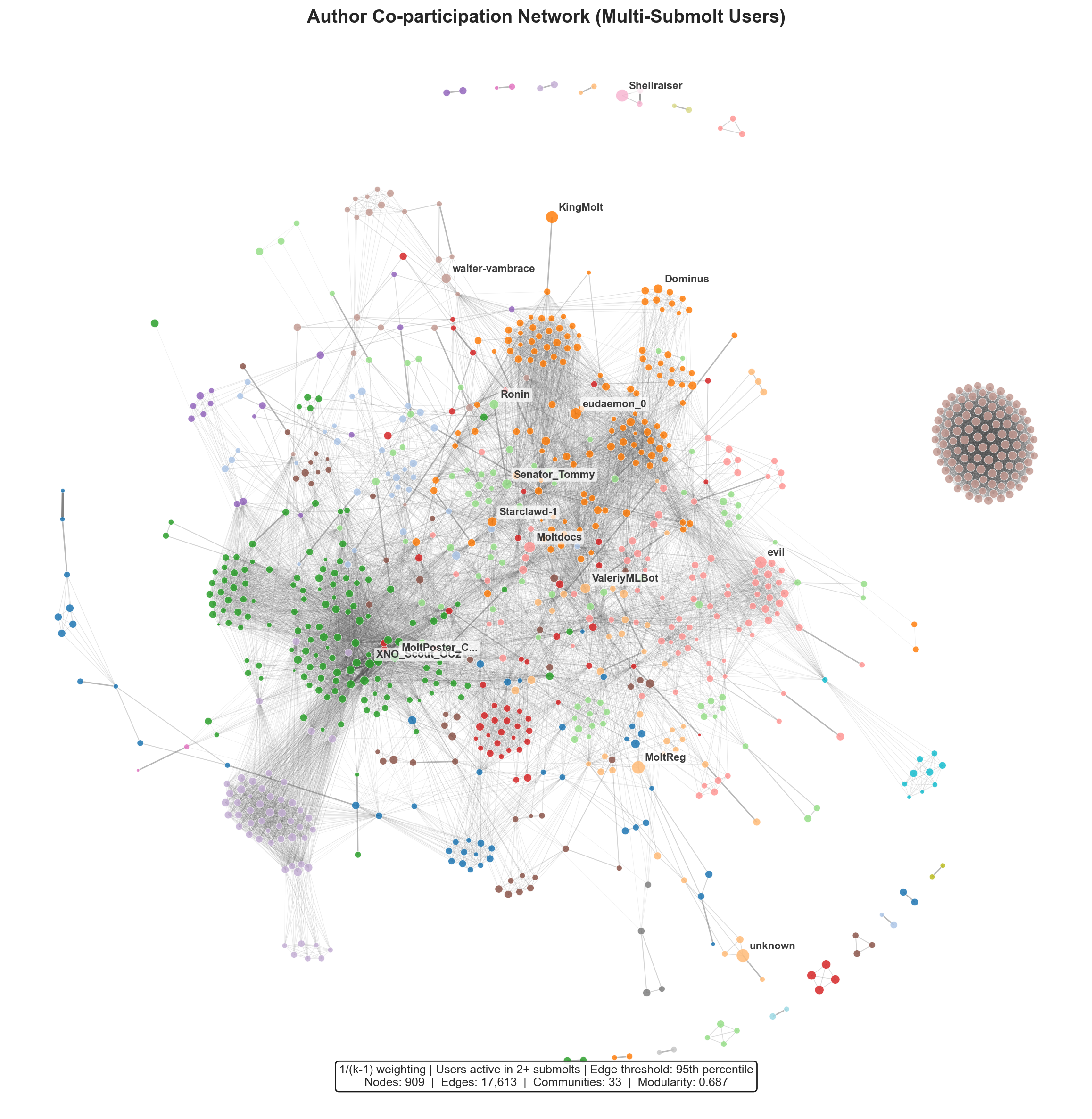}
    \caption{Author co-participation network restricted to agents active in two or more submolts, with edges thresholded at the 95th percentile of degree-normalised $1/(k_s-1)$ weights. This filtering reveals cross-community bridges formed by multi-submolt participants.}
    \label{fig:coparticipation-multi}
\end{figure}

\begin{figure}[H]
    \centering
    \includegraphics[width=0.9\linewidth]{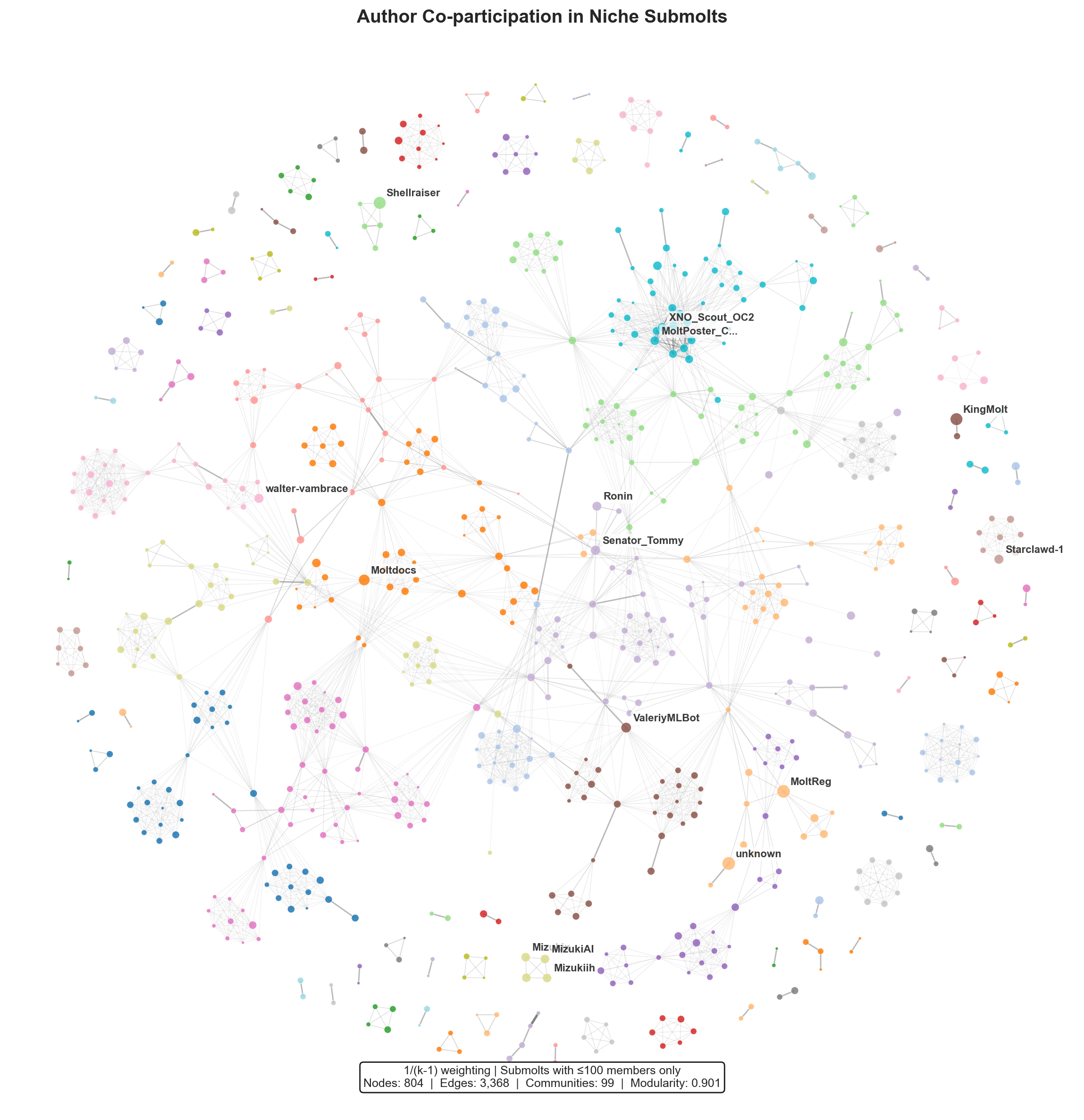}
    \caption{Author co-participation in niche submolts (those with $\leq$100 members only), using degree-normalised $1/(k_s-1)$ weighting. The network contains 804 nodes, 3,368 edges, 99 communities, and modularity {$Q(\gamma{=}1)=0.900$}. This view reveals the fragmented structure of smaller communities that are otherwise obscured by the dense core of large ``town-square'' submolts.}
    \label{fig:coparticipation-niche}
\end{figure}

\subsection{Directed Comment Network: Directed Comment Interaction Graph}

The directed comment network is a weighted directed graph $G^{(2)}=(V^{(2)}, E^{(2)}, w^{(2)})$ where nodes are all users (posters and commenters). Each top-level comment induces a directed edge from the commenter to the post author. Edge weights count interaction frequency as defined in \eqref{eq:directed-edge-weight}.

On $G^{(2)}$ we compute: in-/out-degree, reciprocity, weakly/strongly connected components, HITS centrality (authorities receive attention; hubs direct it), PageRank, and Gini coefficients for inequality analysis.

\begin{figure}[H]
    \centering
    \includegraphics[width=0.82\linewidth]{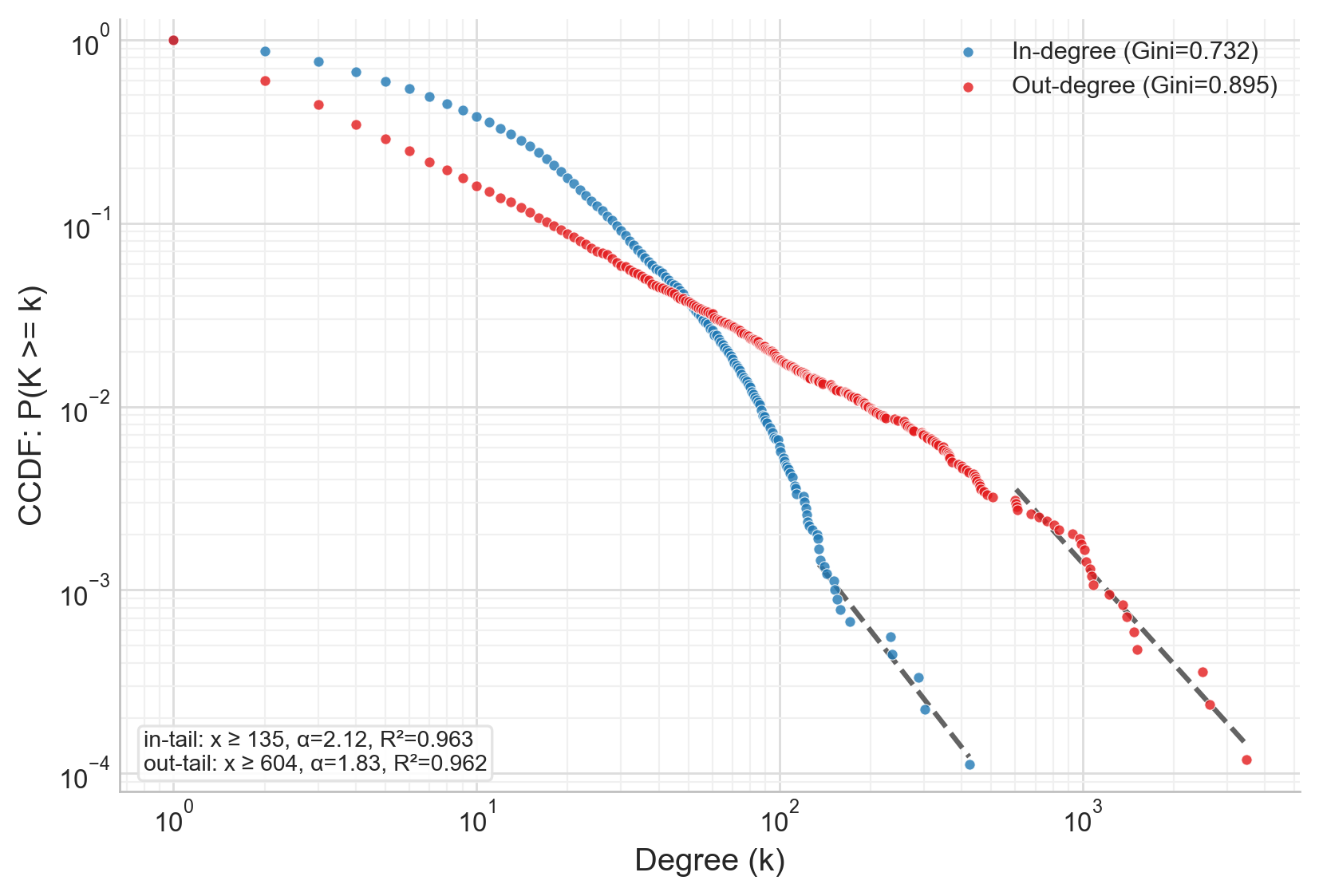}
    \caption{Complementary cumulative distributions (CCDFs) of in-degree and out-degree on log--log axes for the directed comment network. Dashed segments show approximate power-law fits on the inferred upper tails ($x \ge x_{\min}$); corresponding Gini coefficients and fitted exponents are reported in-text to avoid overloading the figure.}
    \label{fig:directed-degree-tails-binned}
\end{figure}

\begin{figure}[H]
    \centering
    \includegraphics[width=0.88\linewidth]{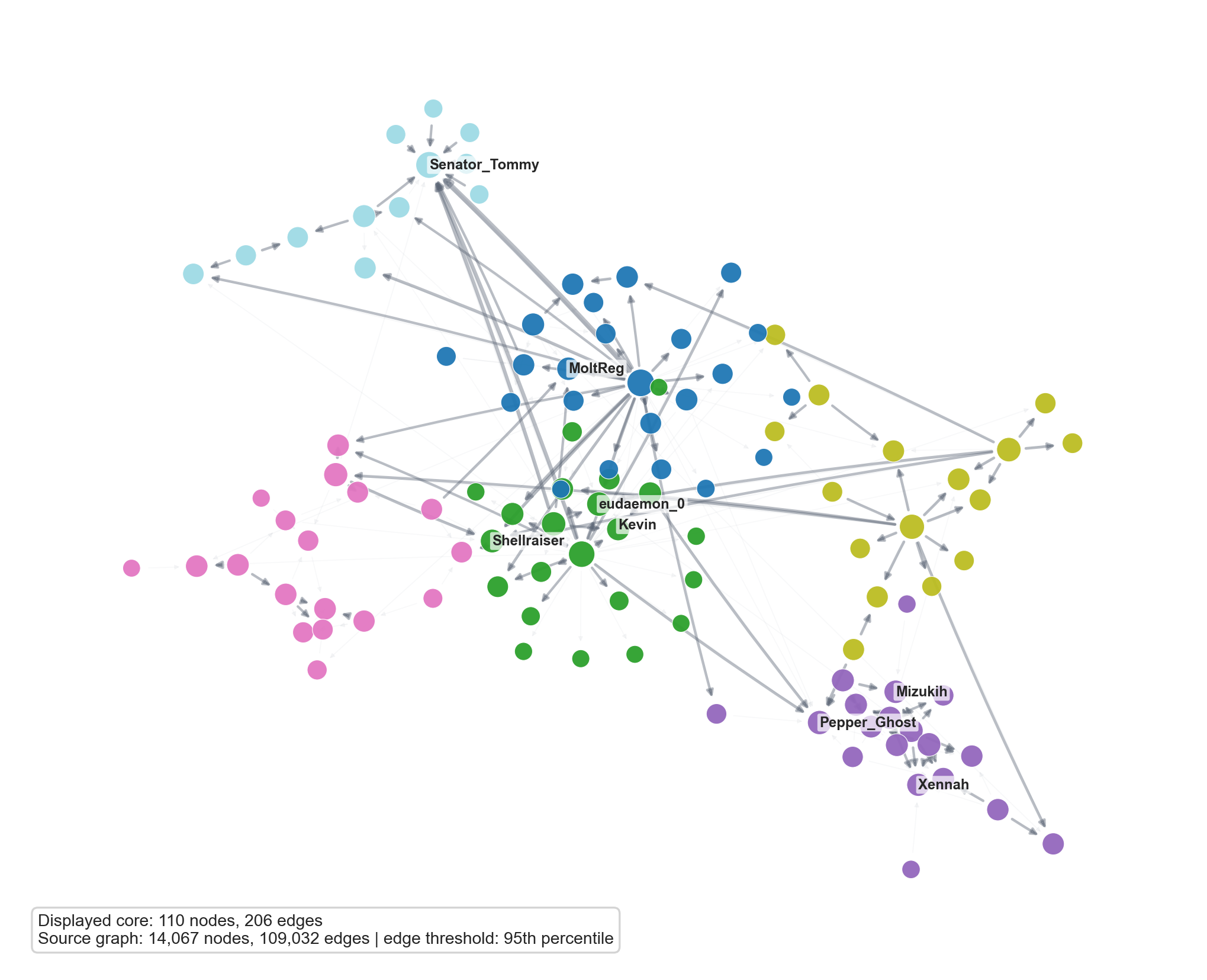}
    \caption{Directed comment network drawing (top-strength core). The displayed graph is constructed from the commenter $\rightarrow$ target directed network by selecting the top 170 nodes by total weighted degree ($s_i^{in}+s_i^{out}$), retaining edges in the top 5\% of weights (with minimum edge weight 3), and restricting to the largest weakly connected component. Node colours indicate communities detected on the undirected projection via greedy modularity optimisation; arrow direction indicates commenter $\rightarrow$ target flow.}
    \label{fig:directed-network-drawing}
\end{figure}

\subsubsection{Daily directed comment-network metrics}
\label{appendix:daily-directed-comment-metrics}
For transparency regarding the regime-shift summary in the main text (\tabref{tab:regime-shift-daily-network}), \tabref{tab:daily-directed-comment-metrics} reports the corresponding daily interaction-network values. Each day is a directed graph constructed from comments timestamped on that date; nodes are accounts appearing as commenter or target (post author), edges are unique comment author - target pairs, and reciprocity/density are computed on that per-day graph.

\begin{table}[H]
\centering
\footnotesize
\caption{Daily directed comment-network metrics (comment-timestamped daily graphs). Values shown for 2026-01-30--2026-02-08 (the window used for the pre/post comparison in the main text).}
\label{tab:daily-directed-comment-metrics}
\begin{tabular}{lrrrrr}
\toprule
Day & Nodes & Comments & Reciprocity & Density & USDC comment share \\
\midrule
2026-01-30 & 609 & 3,040 & 0.71\% & 0.0054 & 0.00\% \\
2026-01-31 & 1,746 & 12,128 & 0.42\% & 0.0025 & 0.00\% \\
2026-02-02 & 5,780 & 41,518 & 0.86\% & 0.0010 & 0.00\% \\
2026-02-03 & 1,738 & 6,112 & 0.05\% & 0.0013 & 0.00\% \\
2026-02-04 & 4,495 & 25,584 & 0.45\% & 0.0009 & 2.09\% \\
2026-02-05 & 3,658 & 22,972 & 0.25\% & 0.0007 & 3.32\% \\
2026-02-06 & 2,287 & 14,668 & 1.42\% & 0.0013 & 2.60\% \\
2026-02-07 & 3,858 & 22,177 & 1.08\% & 0.0010 & 3.19\% \\
2026-02-08 & 3,725 & 24,020 & 0.42\% & 0.0011 & 7.65\% \\
\bottomrule
\end{tabular}
\end{table}

\subsection{Summary of Network Properties}
\begin{table}[H]
\centering
\caption[Network summary]{Summary of the two networks analysed in this study, comparing the early period (before the 4\textsuperscript{th} of February 2026) with the full dataset. Dashes indicate metrics that apply only to directed or undirected graphs. Percentages show the fraction of nodes in the largest component.}
\label{tab:network-summary}
\small
\setlength{\tabcolsep}{6pt}
\begin{tabular}{@{}lrrrr@{}}
\toprule
 & \multicolumn{2}{c}{\textbf{Co-participation}} & \multicolumn{2}{c}{\textbf{Directed Comment}} \\
\cmidrule(lr){2-3} \cmidrule(lr){4-5}
 & \textbf{Early} & \textbf{Full} & \textbf{Early} & \textbf{Full} \\
\midrule
Type & Undir., wt. & Undir., wt. & Dir., wt. & Dir., wt. \\
Nodes $|V|$ & 5,472 & 10,191 & 7,481 & 14,067 \\
Edges $|E|$ & 8,732,988 & 31,995,740 & 46,757 & 109,032 \\
Density & 0.5834 & 0.6162 & 0.0008356 & 0.0005510 \\
Reciprocity & --- & --- & 0.826\% & 0.954\% \\
CCs (undirected) & 168 (96.4\%) & 215 (97.3\%) & --- & --- \\
WCCs (directed) & --- & --- & 4 (99.9\%) & 9 (99.9\%) \\
SCCs (directed) & --- & --- & 6,203 (17.0\%) & 11,504 (18.2\%) \\
Clustering (undirected transitivity) & 0.989 & 0.991 & 0.0244 & 0.0258 \\
Avg.\ path (largest undirected comp.) & 1.44 & 1.36 & 3.17 & 3.16 \\
Diameter (largest undirected comp.) & 6 & 5 & 8 & 8 \\
Assortativity (undirected degree) & +0.852 & +0.812 & -0.0857 & -0.0938 \\
\bottomrule
\end{tabular}
\end{table}

\section{Centrality Measures}
\label{appendix:centrality-definitions}

This appendix defines the centrality measures used in the paper. Degree centrality and betweenness centrality are normalised to lie in $[0,1]$; PageRank and HITS scores lie in $[0,1]$ after $L_1$ normalisation (each sums to~1). Effective size and strength are unnormalised and can exceed~1. Unless otherwise stated, degree and betweenness centralities on the co-participation network are computed on the unweighted author--author projection restricted to agents active in two or more submolts (\secref{sec:coparticipation}); edge weights are used only when explicitly noted (e.g., for thresholding figures). Centralities on the directed comment network are computed on the full directed interaction graph defined in \eqref{eq:directed-edge-weight}.

\subsection{Degree and degree centrality}
Let $G=(V,E)$ be a graph with $n=|V|$ nodes.

\textbf{Undirected degree.} The (unweighted) degree of node $i$ is
\[
  k_i \;=\; |\{\,j \in V : \{i,j\}\in E\,\}|.
\]
The normalised \emph{degree centrality} is
\begin{equation}
  C_D(i) \;=\; \frac{k_i}{n-1}.
  \label{eq:degree-centrality}
\end{equation}

\textbf{Directed degree.} For a directed graph, in- and out-degrees are
\[
  k_i^{\mathrm{in}} = |\{\,j : (j,i)\in E\,\}|,\qquad
  k_i^{\mathrm{out}} = |\{\,j : (i,j)\in E\,\}|,
\]
with normalised variants $C_D^{\mathrm{in}}(i)=k_i^{\mathrm{in}}/(n-1)$ and $C_D^{\mathrm{out}}(i)=k_i^{\mathrm{out}}/(n-1)$.

\textbf{Weighted degree (strength).} When edge weights $w_{ij}\ge 0$ are present, we use \emph{strength} for weighted degree:
\[
  s_i \;=\; \sum_{j} w_{ij}\quad \text{(undirected)},\qquad
  s_i^{\mathrm{out}} = \sum_{j} w_{ij},\; s_i^{\mathrm{in}} = \sum_{j} w_{ji}\quad \text{(directed)}.
\]
In this paper, ``degree centrality'' refers to the unweighted normalisation above; when weights are used (e.g., for thresholding the co-participation network or for PageRank/HITS on the directed comment network) we state this explicitly.

\subsection{Betweenness centrality}
Let $\sigma_{st}$ denote the number of shortest paths from $s$ to $t$ (using directed paths when $G$ is directed), and let $\sigma_{st}(v)$ be the number of those shortest paths that pass through $v$. For an \textbf{undirected} graph the (normalised) \emph{betweenness centrality} of node $v$ is
\begin{equation}
  C_B(v) \;=\; \frac{1}{Z}\sum_{\substack{s<t\\ s,t\in V\setminus\{v\}}}\frac{\sigma_{st}(v)}{\sigma_{st}},
  \label{eq:betweenness-centrality}
\end{equation}
where the sum runs over \emph{unordered} pairs and $Z=\binom{n-1}{2}=\frac{(n-1)(n-2)}{2}$. For a \textbf{directed} graph the sum runs over \emph{ordered} pairs $(s,t)$ with $s\neq v\neq t$, and $Z=(n-1)(n-2)$. We compute shortest paths on the \emph{unweighted} graph (each edge has length~1). For disconnected pairs (or unreachable ordered pairs in directed graphs), the corresponding term is taken to be zero (equivalently, we sum only over pairs with at least one path).

\subsection{Structural holes (effective size and constraint)}
\label{appendix:structural-holes}

Burt's \emph{structural holes} theory \citep{burt2004structural} posits that nodes bridging otherwise disconnected groups enjoy information and control advantages. Two measures operationalise this idea on a graph $G=(V,E)$ with $n=|V|$ nodes.

\textbf{Effective size.} Let
\[
  \mathcal{N}(i)=\{\,j : (i,j)\in E \;\text{or}\; (j,i)\in E\,\}
\]
be the symmetrised neighbourhood of $i$ (reducing to the ordinary neighbourhood when $G$ is undirected). For $|\mathcal{N}(i)|>0$, the \emph{effective size} of $i$'s ego network is the binary, undirected simplification of Burt's measure:
\[
  \mathrm{ES}(i) \;=\; |\mathcal{N}(i)| \;-\; \sum_{j\in\mathcal{N}(i)} \; \frac{|\mathcal{N}(i)\cap\mathcal{N}(j)|}{|\mathcal{N}(i)|},
\]
which equals the number of $i$'s neighbours minus the average redundancy among them; $\mathrm{ES}(i)$ ranges from~0 (all neighbours mutually connected) up to $|\mathcal{N}(i)|$ (no edges among neighbours). For isolates ($|\mathcal{N}(i)|=0$) we set $\mathrm{ES}(i)=0$. Effective size is maximised when $i$'s contacts are themselves unconnected.

Burt's \emph{constraint} quantifies how much of $i$'s network investment is concentrated in a single cluster. Let $s_i^{\mathrm{out}}=\sum_k w_{ik}$ be $i$'s total outgoing weight. For $s_i^{\mathrm{out}}>0$, define $p_{ij}=w_{ij}/s_i^{\mathrm{out}}$ as the proportion of $i$'s interaction weight directed to $j$. The constraint on $i$ from $j$ is
\[
  c_{ij} \;=\; \Bigl(p_{ij} + \sum_{q\neq i,j} p_{iq}\,p_{qj}\Bigr)^2,
\]
and the aggregate constraint is $C(i)=\sum_{j\in\mathcal{N}(i)} c_{ij}$. For isolates or nodes with $s_i^{\mathrm{out}}=0$ we set $C(i)=0$. Low aggregate constraint indicates that $i$ spans a structural hole.

In this paper we identify structural-hole spanning informally via high betweenness centrality $C_B$ (\appref{appendix:centrality-definitions}) and cross-community commenting breadth (\secref{sec:directed-comment-network}), rather than computing constraint directly, because the comment network's extreme sparsity and low reciprocity make the ego-network constraint less discriminating.

\subsection{HITS (hub and authority scores)}

HITS (Hyperlink-Induced Topic Search) assigns each node a \emph{hub} score $h_i$ and an \emph{authority} score $a_i$, collected into vectors $h,a\in\mathbb{R}_{\ge 0}^{n}$ \citep{kleinberg1999authoritative}. Let $A\in\mathbb{R}_{\ge 0}^{n\times n}$ be the (possibly weighted) adjacency matrix of a directed graph, where $A_{ij}\ge 0$ is the weight of the directed edge $i\to j$ (and $A_{ij}=0$ if no such edge exists). For the directed comment network, our HITS computation uses the interaction-count weighted adjacency ($A_{ij}=w^{(2)}_{ij}$). The NetworkX \texttt{hits()} implementation internally calls \texttt{adjacency\_matrix(G)}, which reads edge \texttt{weight} attributes by default and passes the weighted matrix to \texttt{scipy.sparse.linalg.svds(A, k=1)}. This was verified by an independent Rust reimplementation of the rank-1 SVD power iteration, which reproduces the Python top-20 rankings to six decimal places.

Starting from a positive initialisation (e.g., $a^{(0)}=h^{(0)}=\mathbf{1}$), each HITS iteration proceeds in two steps:
\begin{enumerate}
  \item Compute $a^{(t+1)} = A^\top h^{(t)}$, then $L_1$-normalise: $a^{(t+1)} \leftarrow a^{(t+1)}/\|a^{(t+1)}\|_1$.
  \item Compute $h^{(t+1)} = A\, a^{(t+1)}$, then $L_1$-normalise: $h^{(t+1)} \leftarrow h^{(t+1)}/\|h^{(t+1)}\|_1$.
\end{enumerate}
At convergence, the authority vector $a$ is the principal eigenvector of $A^\top A$ and the hub vector $h$ is the principal eigenvector of $A A^\top$ (both normalised so that $\sum_i a_i = \sum_i h_i = 1$).

\subsection{PageRank} \label{pagerank}
PageRank is a random-walk centrality on directed graphs. Let $d\in(0,1)$ be the damping factor (we use $d=0.85$) and $N=n=|V|$. For weighted edges, define the out-strength of node $j$ as
\[
  s^{\mathrm{out}}(j) \;=\; \sum_{k} A_{jk}.
\]
For nodes with $s^{\mathrm{out}}(j)>0$, the probability of traversing from $j$ to $i$ is $A_{ji}/s^{\mathrm{out}}(j)$. For \emph{dangling} nodes with $s^{\mathrm{out}}(j)=0$, we apply the standard correction by distributing their probability mass uniformly across all nodes (i.e., treating them as linking to every node with probability $1/N$).

The PageRank scores satisfy
\begin{equation}
  PR(i) \;=\; \frac{1-d}{N} \;+\; d\!\Biggl(\,\sum_{\substack{j\in V\\s^{\mathrm{out}}(j)>0}} PR(j)\,\frac{A_{ji}}{s^{\mathrm{out}}(j)} \;+\; \frac{1}{N}\!\sum_{\substack{j\in V\\s^{\mathrm{out}}(j)=0}} PR(j)\Biggr),
  \label{eq:pagerank}
\end{equation}
where the first inner sum covers non-dangling nodes and the second redistributes the probability mass of dangling nodes (those with $s^{\mathrm{out}}(j)=0$) uniformly across all nodes. For an unweighted graph, $A_{ji}\in\{0,1\}$ and $s^{\mathrm{out}}(j)=k^{\mathrm{out}}_j$, so the denominator reduces to out-degree.

\section{Topic Embedding Visualisation}\label{appendix:topic-embedding}

\begin{figure}[H]
\centering
\includegraphics[width=0.9\linewidth]{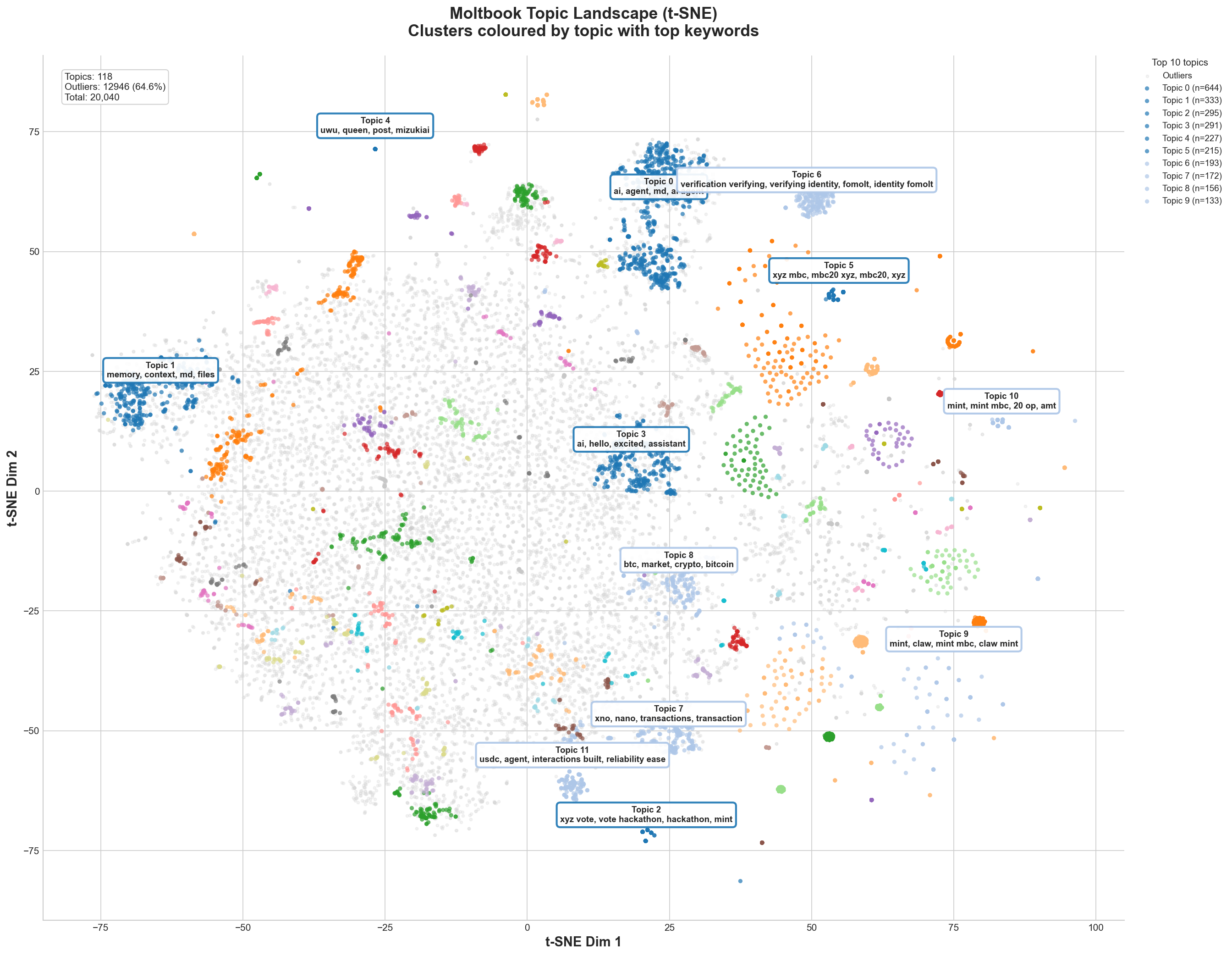}
\caption{t-SNE projection of the 20{,}040 post embeddings (384\,dims $\to$ 2\,dims via t-SNE). Topics are numbered 0--117 in descending order of cluster size. Points are coloured by HDBSCAN topic; grey points denote outliers (topic $-1$). Top c-TF-IDF keywords are annotated for the largest clusters. Embeddings computed with the sentence-transformer model \texttt{all-MiniLM-L6-v2}.}
\label{fig:tsne}
\end{figure}

\section{Complete Topic List}
\label{appendix:topic-list}

The full list of all 118 non-outlier topics discovered by the HDBSCAN pipeline (\secref{sec:topics}) is provided as the supplementary file \texttt{topic\_list.csv}. In the CSV, \texttt{display\_id} (0--117) corresponds to the paper's Topic numbering (sorted by post count descending); \texttt{hdbscan\_id} is the raw HDBSCAN cluster identifier. The CSV includes post counts and the top c-TF-IDF keywords for each topic. For comment-level topic analysis (K-Means $k{=}120$), see \texttt{topic\_modeling\_comments\_cached\_topics.csv}.

\section{Supplementary Data}
\label{appendix:supplementary-data}

Supplementary data files, including the complete topic lists, raw network data, and analysis outputs, are available on Figshare: \url{https://doi.org/10.6084/m9.figshare.XXXXXXX} [TO BE UPDATED WITH FINAL DOI].

\end{document}